%% file: MAIN.tex
\documentclass[11pt]{article}
\usepackage{fullpage}
\usepackage{amsmath,amssymb}
\usepackage{enumitem}
\usepackage{algorithm}
\usepackage[noend]{algpseudocode}
\usepackage{color}
\usepackage{setspace}
\usepackage{wrapfig}
\usepackage{graphicx}
\usepackage{multicol}
\usepackage{chngcntr}
\usepackage{apptools}

\usepackage[polutonikogreek,english]{babel}
\usepackage[utf8x]{inputenx}
\newcommand{\greek}[1]{{\selectlanguage{polutonikogreek}#1}}

\newenvironment{sketch-proof}{%
  \proof}{\hfill{\small$\Box$}\endproof}

\algrenewcommand\alglinenumber[1]{\tiny #1:}

\newcommand{\vq}[1]{\textsc{qv\scriptsize{(#1)}}} 
\newcommand{\tg}[1]{tg_{#1}}

\graphicspath{ {sim_charts/} }

\input{macros.tex}

\input{nn_macros.tex}
\newcommand{\ch}[1]{\textcolor{magenta}{#1}}

\newcommand{\ppp}[1]{\smallskip\noindent{\bf #1}}

\newcommand{\osam}{{\sc OhSam}}
\newcommand{\omam}{{\sc OhMam}}
\newcommand{\algsw}{{\sc Erato}}
\newcommand{\algmw}{{\sc Erato-mw}}
\renewcommand{\ABD}{{\sc ABD}}
\newcommand{\ABDmwmr}{{\sc ABD-mw}}
\newcommand{\swimp}{{\sf LB}}

\newcommand{\sliq}{{\sc Sliq}}

\newcommand{\metrics}{{exchanges}}

\begin{document}

\title{Unleashing and Speeding Up Readers\\
in Atomic Object Implementations}

\author{	Chryssis Georgiou
			\thanks{University of Cyprus, Department of Computer Science, Nicosia, Cyprus. Email: {\tt chryssis@cs.ucy.ac.cy}}
			\and
			Theophanis Hadjistasi
		 	\thanks{University of Connecticut, Storrs CT, USA. Email: {\tt theo@uconn.edu, aas@uconn.edu}}
			\and 
			Nicolas Nicolaou
			\thanks{KIOS Research Center of Excellence, University of Cyprus, Nicosia, Cyprus Email: {\tt nicolasn@cs.ucy.ac.cy} }
			\and 
			Alexander Schwarzmann$^*$}

\maketitle

\vspace{-4mm} 

\begin{abstract} 
Providing efficient emulations of atomic read/write objects 
in asynchronous, crash-prone, message-passing systems
is an important problem in distributed computing. 
Communication latency is a factor that typically dominates
the performance of message-passing systems, consequently 
the efficiency of algorithms implementing atomic objects 
is measured in terms of the number of communication exchanges
involved in each read and write operation.
The seminal result of Attiya, Bar-Noy, and Dolev established
that two pairs of communication exchanges, or equivalently
two round-trip communications, are sufficient.
Subsequent research examined the possibility of
implementations that involve less than four exchanges.
The work of Dutta et al.~showed that for single-writer/multiple-reader
({\small SWMR}) settings two exchanges are sufficient, provided that
the number of readers is severely constrained with respect to
the number of object replicas in the system and the number
of replica failures, and also showed that no two-exchange implementations of 
multiple-writer/multiple-reader ({\small MWMR}) objects are possible.
Later research focused on providing implementations
that remove the constraint on the number of readers, while having
read and write operations that use variable number of communication
exchanges, specifically two, three, or four exchanges.

This work presents two advances in the state-of-the-art in this area.
Specifically, for {\small SWMR} and {\small MWMR} systems
algorithms are given in which read operations take \emph{two or three} exchanges.
This improves on prior works where
read operations took either $(a)$ three exchanges, or 
$(b)$ two or four exchanges.
The number of readers in the new algorithms is \emph{unconstrained},
and  write operations take the same number of exchanges
as in prior work (two for {\small SWMR} and four for {\small MWMR} settings).
The correctness of algorithms is rigorously argued.
The paper presents an empirical  study using the  NS3 simulator
that compares the performance of relevant algorithms,
demonstrates the practicality of the new algorithms, and
identifies settings in which their performance is clearly superior.
\end{abstract}


\thispagestyle{empty}

\newpage

\section{Introduction}
\label{sec:erato:intro}
\input{introduction.tex}

\section{Models and Definitions}
\label{sec:erato:model}
\input{model.tex}

\section{SWMR Algorithm \algsw{}}
\label{sec:erato:sw}
\input{erato-SW.tex}

\section{MWMR Algorithm \algmw{}}
\label{sec:erato:mw}
\input{erato-MW.tex}

\section{Empirical Evaluations}
\label{sec:erato:simulations}
\input{simulation.tex}

\section{Conclusions}
\label{sec:erato:conclude}

We focused on the problem of emulating atomic read/write shared objects 
in the asynchronous, crash-prone, message-passing settings
with the goal of synthesizing algorithms where read operations
can \emph{always} terminate in \emph{less} than two communication
round-trips. 
%
We presented such algorithms for the {\small SWMR} and {\small MWMR} models.
We rigorously reasoned about the correctness of our algorithms. 
The algorithms impose no constraints on the number of readers, 
and no constraints on the number of
 writers (in the {\small MWMR} model). 
The algorithms are shown to be optimal in terms of \emph{communication $\metrics$} 
with unconstrained participation.
The empirical study demonstrates the practicality of the new algorithms,
and identifies
settings in which their performance is clearly superior.


\bibliographystyle{acm}
\bibliography{biblio}

\end{document}

%% file: macros.tex
\newtheorem{theorem}{Theorem}
\newtheorem{lemma}[theorem]{Lemma}

\newcommand{\prf}[1]{{}}

\newenvironment{Proof}{\noindent{\bf Proof.}}{\hfill$\Box$\FF}

\newtheorem{Def}{Definition}[section]

\newcommand{\sacode}[5]
{ \vspace{.06in} \hrule \vspace{.06in} 
 \noindent {\bf #1}: \\
 \footnotesize \noindent {\bf Signature:}\B \nobreak
 \normalsize \begin{quote} \nobreak #2 \end{quote}
 \footnotesize \noindent {\bf States:}\B \nobreak
 \begin{quote} \nobreak #3 \end{quote}
 \noindent {\bf Transitions:} \nobreak
 \vspace{-.2in} \nobreak
 \normalsize #4
 \vspace{-.06in} \hrule \vspace{.06in} 
}

\newcommand{\act}[1]{%
    \relax\ifmmode
        \mathord{\mathcode`\-="702D\sf #1\mathcode`\-="2200}%
    \else
        $\mathord{\mathcode`\-="702D\sf #1\mathcode`\-="2200}$%
    \fi
}

\newcommand{\tup}[1]{%
    \relax\ifmmode
      \langle #1 \rangle%
    \else
        $\langle$#1$\rangle$%
    \fi
}

\newcommand{\seq}[1]{%
    \relax\ifmmode
      \langle \! \langle #1 \rangle \! \rangle%
    \else
        $\langle \! \langle$ #1 $\rangle \! \rangle$%
    \fi
}

\newcommand{\B}{\vspace*{-\smallskipamount}}

\newcommand{\FF}{\vspace*{\medskipamount}}

\newcommand{\ms}[1]{%
    \relax\ifmmode
        \mathord{\mathcode`\-="702D\it #1\mathcode`\-="2200}%
    \else
        {\it #1}%
    \fi
}

\newcommand{\lit}[1]{%
    \relax\ifmmode
        \mathord{\mathcode`\-="702D\sf #1\mathcode`\-="2200}%
    \else
        {\it #1}%
    \fi
}

\newcommand{\XDK}[1]{}
\newcommand{\remove}[1]{} 
\newcommand{\uselater}[1]{} 

\renewcommand{\setminus}{-}



%% file: nn_macros.tex
\providecommand{\tup}[1]{%
    \relax\ifmmode
      \langle #1 \rangle%
    \else
        $\langle$#1$\rangle$%
    \fi
}

\renewcommand{\act}[1]{%
    \relax\ifmmode
        \mathord{\mathcode`\-="702D\sf #1\mathcode`\-="2200}%
    \else
        $\mathord{\mathcode`\-="702D\sf #1\mathcode`\-="2200}$%
    \fi
}

\makeatletter
\def\mainlistofsymbols{
  \normalsize
  \vspace*{1.5 em}
  \@starttoc{los}
}

\def\partonelistofsymbols{
  \normalsize
  \vspace*{1.5 em}
  \@starttoc{p1los}
}

\def\parttwolistofsymbols{
  \normalsize
  \vspace*{1.5 em}
  \@starttoc{p2los}
}

\def\l@symbol#1#2{\addpenalty{-\@highpenalty} \vskip 4pt plus 2pt
{\@dottedtocline{0}{0em}{8em}{#1}{#2}}}
\makeatother

\newcommand{\newhiddensym}[2]{%
}

\newcommand{\algIOA}[2]{\ifmmode{\text{#1}_{#2}}\else{$\text{#1}_{#2}$}\fi}

\newcommand{\EX}{\ifmmode{\xi}\else{$\xi$}\fi}
\newcommand{\EXF}{\ifmmode{\phi}\else{$\phi$}\fi}

\newcommand{\inter}[1]{
	\ifmmode{\left(\bigcap_{\mathcal{Q}\in#1}\mathcal{Q}\right)}
	\else{$\left(\bigcap_{\mathcal{Q}\in#1}\mathcal{Q}\right)$}
	\fi
}

\newcommand{\idSet}{\mathcal{I}}
\newcommand{\wSet}{\mathcal{W}}
\newcommand{\rdSet}{\mathcal{R}}
\newcommand{\srvSet}{\mathcal{S}}

\newcommand{\op}{\pi}

\newcommand{\rd}{\rho}
\newcommand{\wrt}{\omega}

\mathchardef\mhyphen="2D

\newcommand{\pr}{p}
\newcommand{\rdr}{r}

\newcommand{\srvr}{s}

\newcommand{\bef}{\rightarrow}

\newcommand{\vid}[1]{\ifmmode{\nu_{#1}}\else{$\nu_{#1}$}\fi}

\newcommand{\seen}{\ifmmode{seen}\else{$seen$}\fi}

\newcommand{\cwfr}{{\sc CwFr}}

\newcommand{\ABD}{{\sc ABD}}

\newcommand{\maxts}[1]{\ifmmode{maxTS_{#1}}\else{$maxTS_{#1}$}\fi}
\newcommand{\maxtag}[1]{\ifmmode{maxTag_{#1}}\else{$maxTag_{#1}$}\fi}
\newcommand{\maxpair}[1]{\ifmmode{maxMPair_{#1}}\else{$maxMPair_{#1}$}\fi}
\newcommand{\mintag}[1]{\ifmmode{minTag_{#1}}\else{$minTag_{#1}$}\fi}
\newcommand{\maxps}{\ifmmode{maxPS}\else{$maxPS$}\fi}
\newcommand{\conftg}[1]{\ifmmode{confirmed_{#1}}\else{$confirmed_{#1}$}\fi}
\newcommand{\maxconftag}{\ifmmode{\ms{maxCT}}\else{$maxCT$}\fi}

%% file: introduction.tex
Emulating atomic \cite{Lamport79} (or linearizable \cite{HW90}) read/write objects in 
asynchronous, 
message-passing environments with crash-prone processors 
is a fundamental problem in distributed computing. 
%
%
To cope with processor failures, distributed object 
implementations
use \emph{redundancy} by replicating the object 
at multiple network locations.
%
%
Replication masks failures, however it introduces the problem 
of consistency
because operations may access different object replicas
possibly containing obsolete values.
Atomicity is the 
most intuitive consistency semantic as it provides the illusion of a single-copy object 
that serializes all accesses such that each read operation returns the value of the latest
preceding write operation.\smallskip

\noindent{\bf Background and Prior Work.}
The seminal work of Attiya, Bar-Noy, and Dolev \cite{ABD96}
provided an algorithm, colloquially referred to as \ABD{}, 
that implements 
{\small SWMR} (Single Writer, Multiple Reader)
atomic objects
in message-passing crash-prone asynchronous environments. 
Operations are ordered using
logical \emph{timestamps} associated with each value. 
Operations terminate provided some majority of replica servers does not crash. 
Writes involve a single communication round-trip involving 
\emph{two} communication exchanges. Each read operation 
 takes two rounds involving in  \emph{four} communication exchanges.
%
%
%
%
Subsequently, Lynch et al.~\cite{LS97} showed how to implement 
{\small MWMR} (Multi-Writer, Multi-Reader)
atomic memory where both read and write operations take two communication
round trips, for a total of \emph{four} exchanges.

%
Dutta et al.~\cite{CDGL04} introduced a \emph{fast} {\small SWMR} implementation
where both reads and writes involve 
\emph{two} exchanges (such operations are called `fast').
It was shown that this is possible only when 
the number of readers $r$ is constrained with respect to the number of servers $s$
and the number of server failures $f$, viz. $r<\frac{s}{f}-2$.
%
%
Other works
focused on relaxing 
the bound on the number of readers in the service by proposing 
hybrid approaches where some operations complete 
in \emph{one} and others in \emph{two} rounds, e.g.,
 \cite{EGMNS09}.
%

Georgiou et al.~\cite{GNS08}
introduced Quorum Views,
client-side tools that examine the distribution of
the latest value among the replicas in order to enable fast
read operations (two exchanges).
%
A {\small SWMR} algorithm,
called \sliq{}, was given that requires at
least one single slow read per write operation, and where all
writes are fast. 
%
A later work \cite{GNRS11} generalized the client-side decision tools
and presented a {\small MWMR} algorithm, 
called \cwfr{}, that allows fast read operations. 
%

Previous works considered only client-server communication round-trips. 
Recently,   
Hadjistasi
et al.~\cite{HNS17} showed
that atomic operations do not necessarily require complete communication
round trips, by introducing server-to-server communication.
They presented a {\small SWMR} algorithm, called \osam{}, where
reads take \emph{three} exchanges:
{two of these are between clients and servers, and one is among servers;
%
their {\small MWMR} algorithm, called \omam{}, 
uses a similar approach.
%
These algorithms do not impose constrains on reader participation
and perform a  modest amount of local computation,
resulting in negligible computation overhead.

\begin{table}[t!h]
\begin{center}
{\small
\begin{tabular}{| c || l | c | c | c | c |}
\hline
  Model & Algorithm & Read Exch. & Write Exch. & Read Comm. & Write Comm.\\
  \hline\hline
  {\sc swmr} & \ABD{} \cite{ABD96} & 4 & 2 & $4|\srvSet|$ & $2|\srvSet|$\\ 
  \hline 
  {\sc swmr}  & \osam{} \cite{HNS17} & 3 & 2 & $|\srvSet|^{2} + 2|\srvSet|$ & $2|\srvSet|$\\
  \hline 
  {\sc swmr}  & \sliq{} \cite{GNS08} & 2 or 4 & 2 & $4|\srvSet|$ & $2|\srvSet|$\\ 
  \hline
  {\sc swmr}  & \algsw{}  & 2 or 3 & 2 & $|\srvSet|^{2} + 3|\srvSet|$ & $2|\srvSet|$\\
  \hline\hline
  {\sc mwmr}  & \ABDmwmr{} \cite{ABD96,LS97} & 4 & 4 & $4|\srvSet|$ & $4|\srvSet|$\\
  \hline
  {\sc mwmr} & \omam{} \cite{HNS17}  & 3 & 4 & $|\srvSet|^{2} + 2|\srvSet|$ & $4|\srvSet|$\\
  \hline
  {\sc mwmr} & \cwfr{} \cite{GNRS11}  & 2 or 4 & 4 & $4|\srvSet|$ & $4|\srvSet|$\\
  \hline
  {\sc mwmr} & \algmw{}  & 2 or 3 & 4 & $|\srvSet|^{2} + 3|\srvSet|$ & $4|\srvSet|$\\
  \hline
\end{tabular}
}
\end{center}
\caption{Summary of communication exchanges and communication complexities.}
\vspace{-\bigskipamount}
\label{erato:table:complexities}
\end{table}

\ppp{Contributions.}
We focus on the gap between one-round and two-round algorithms
by presenting atomic memory algorithms where read operations
can take \emph{at most} ``one and a half rounds,'' i.e.,
complete in either \emph{two} or \emph{three} exchanges.
%
%
Complexity results are shown in Table~\ref{erato:table:complexities}, additional details are as follows.
%

{\bf 1.} We present \algsw{},\footnote{
\greek{Ερατώ}
is a Greek Muse, 
and the authors thank the lovely muse for her inspiration.}
\emph{Efficient Reads for ATomic Objects}, a {\small SWMR} algorithm 
	for atomic objects in the asynchronous
	message-passing model with processor crashes.
	We improve the \emph{three}-exchange read protocol 
	of \osam{} \cite{HNS17}  to allow	
	reads to terminate in either \emph{two} or \emph{three} exchanges
	using client-side tools, Quorum Views, from algorithm \sliq{} \cite{GNS08}.
	%
	During the second exchange,
	based on the distribution of the timestamps,
	the reader may be able to complete the read.
	If not, it awaits for ``enough'' messages from the third exchange
	to complete.
	%
	A key idea
	is that when the reader is ``slow'' it returns the value 
	associated with the \emph{minimum} timestamp, 
	i.e., the value of the previous write that is guaranteed to be complete
        (cf.~\cite{HNS17} and \cite{CDGL04}). 
	Read operations are optimal in terms 
	of exchanges in light of \cite{HNS2017arx}.
	Similarly to \ABD{}, writes take \emph{two}
	exchanges. 
	(Section \ref{sec:erato:sw}.)
	
{\bf 2.}
Using the {\small SWMR} algorithm as the basis,
	we develop a {\small MWMR} algorithm, \algmw{}. 
	 %
The algorithm supports
\emph{three}-exchange read protocol 
based on \cite{HNS17}
	in combination with the iterative technique using quorum views 
	as in  \cite{GNRS11}. 
	%
	Reads take either \emph{two} or 
	\emph{three} exchanges. 
	Writes are similar to \ABDmwmr{} and take
	 \emph{four} communication exchanges (cf.~\cite{LS97}). 
	(Section \ref{sec:erato:mw}.)

{\bf 3.}
We simulate the algorithms using the NS3 simulator and assess their performance
	under practical considerations 
by varying
	the number of participants, frequency of operations, and  network topologies.
	(Section \ref{sec:erato:simulations}.)
	

Improvements in latency 
are obtained in a trade-off for communication complexity. 
Simulation results suggest that in  practical settings, such as  data centers
with well-connected servers,
the communication overhead is not prohibitive.
\vspace{-\medskipamount}

%% file: model.tex
We now present the  model, definitions, and notations used in the paper. 
The system is a collection of crash-prone, {asynchronous processors} 
with unique identifiers (ids).
The ids are from a totally-ordered set $\idSet$
that is composed of three disjoint sets,
set $\wSet$ of writer ids,
set $\rdSet$ of reader ids, and 
set $\srvSet$ of replica server ids.
Each \emph{server}
maintains a copy of the object.

Processors communicate by {exchanging messages} via {asynchronous} point-to-point
reliable channels; messages may be reordered. 
We use the term \emph{broadcast} as a shorthand
denoting sending point-to-point messages to multiple destinations.

A quorum system over a set is a collection of subsets, called quorums, such that
every pair of quorums intersects.
We define a quorum system $\mathbb{Q}$ over the set of 
server ids $\srvSet$ as 
$\mathbb{Q} = \{Q_i : Q_i \subseteq \srvSet \}$;
it follows that for any
$Q_i, Q_j \in \mathbb{Q}$ we have $Q_i \cap Q_i \neq \emptyset$.
We assume that every process in the system is aware of $\mathbb{Q}$.

\ppp{Executions.}
An algorithm $A$ is a collection of
processes, where 
process $A_\pr$ is assigned to processor $\pr\in\idSet$. 
The \emph{state} of processor $\pr$ is determined over a 
set of state variables, and the state of $A$ is a vector that 
contains the state of each process. 
Algorithm $A$ performs a \emph{step}, when some process $p$
(i) receives a message,
(ii) performs local computation,
(iii) sends a message.
Each such action  causes the state at $p$ to change.
An \emph{execution} is an alternating sequence of states and actions 
of $A$ starting with the initial state and ending in a state.

\ppp{Failure Model.}
A process may crash at any point in an execution.
If it crashes, then it stops taking steps; otherwise 
we call the process \emph{correct}.
Any subset of readers and writers may crash.
A quorum $Q \in \mathbb{Q}$ is non-faulty if $\forall \pr \in Q$, $\pr$ is correct.
Otherwise, we say $Q$ is faulty. 
We allow for any server to crash as long one 
quorum is non-faulty.

\ppp{Efficiency and Message Exchanges.} 
Efficiency of implementations is assessed in terms of
\emph{operation latency} and \emph{message complexity}.  
\emph{Latency} of an operation is 
determined by
\emph{computation time} and the \emph{communication delays}.
{Computation time} accounts for all local computation
within an operation.
%
{Communication delays} are measured in terms 
of \emph{communication $\metrics$}.
The protocol implementing each operation involves a collection of
sends (or broadcasts) of typed messages and the corresponding receives.
As defined 
in
\cite{HNS17}, a \emph{communication exchange} 
within an execution of an operation is the set of sends and receives for the 
specific message type.
%
Traditional implementations in the style of \ABD{} \cite{ABD96}
are structured in terms of \emph{rounds},
each consisting of two exchanges, the first, a broadcast, 
is initiated by the process executing an operation, and the second, a convergecast, 
consists of responses to the initiator.
The number of messages that a process expects during a convergecast 
depends on the implementation. 
\emph{Message complexity} measures 
the {worst-case} total number of messages exchanged. 

\ppp{Atomicity.} 
An implementation of a read or a write operation contains an \emph{invocation} action 
and a \emph{response} action.
%
An operation $\op$ is \emph{complete} in an execution, if it contains 
both the invocation and the \emph{matching} response actions for $\op$; otherwise 
$\op$ is \emph{incomplete}. 
An execution is \emph{well formed} if any process invokes one operation at a time. 
We say that an operation $\op$ \emph{precedes} an operation $\op'$ in an execution $\EX$, 
denoted by $\op\bef\op'$, if the response step of $\op$ appears before the invocation step in $\op'$ in $\EX$. 
Two operations are \emph{concurrent} if neither precedes the other. 
The correctness of an atomic read/write object implementation is defined in terms of  
\emph{atomicity} (safety) and \emph{termination} (liveness) properties. 
Termination requires that any operation invoked by a correct process eventually completes.
Atomicity  is defined following \cite{Lynch1996}. For any execution $\EX$, 
if $\Pi$ is the set of all completed read and write operations in $\EX$, 
then there exists a partial order $\prec$ on the operations in $\Pi$, s.t. the following properties are satisfied:
%

\noindent{\bf A1} 
	For any $\op_1,\op_2\in\Pi$ such that $\op_1\bef\op_2$, it cannot be that $\op_2\prec\op_1$

\noindent{\bf A2} 
For any write $\wrt\in\Pi$ and any operation $\op\in\Pi$, then either $\wrt\prec\op$ or $\op\prec\wrt$.

\noindent{\bf A3} 
Every read operation returns the value of the last write preceding it according to $\prec$ (or the initial value if there is no such write)

\ppp{Timestamps and Quorum Views.} 
Atomic object implementations typically use logical timestamps (or tags) 
associated with the written values to impose a partial order on operations
that satisfies the  properties A1, A2, and A3.

\begin{figure}[ht]
\includegraphics[width=\textwidth]{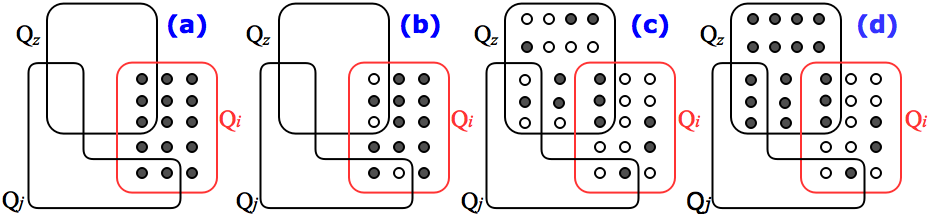}
\caption{\small{(a) \vq{1}, (b) \vq{2}, (c) \vq{3} incomplete write, (d) \vq{3} complete write}}
\label{fig:usup:views}
\end{figure}

A \emph{quorum view} refers to the distribution of the 
highest timestamps that a read operation
witnesses during an exchange. Fig.~\ref{fig:usup:views} illustrates
four different scenarios. Here small circles represent timestamps
received from servers, and dark circles represent the highest
timestamp, and the light ones represent older timestamps.
The quorum system consists of three quorums, $Q_i, Q_j,$ and $Q_z$.
%

Suppose a read  $\rd$ strictly receives values and timestamps from quorum $Q_i$
during an exchange.
%
As presented in \cite{GNRS11}, $\rd$ can distinguish three different 
cases: \vq{1}, \vq{2}, or \vq{3}. Each case can help $\rd$ derive conclusions about the state 
of the latest write (complete, incomplete, unknown).
%
%
If \vq{1} is detected, Fig.~\ref{fig:usup:views}(a),
it means that only one timestamp is received.
This means that the write associated with this timestamp is complete.
If \vq{2} is detected, Fig.~\ref{fig:usup:views}(b), this indicates that the write associated with the 
highest timestamp is still in progress (because older timestamps are detected
in the intersections of quorums).
Lastly, if \vq{3} is detected, the distribution of timestamps does not
provides sufficient information regarding the state of the write. 
This is because there are two possibilities as
shown in Fig.~\ref{fig:usup:views}(c) and \ref{fig:usup:views}(d).
In the former the write is incomplete (no quorum has the highest detected timestamp)
and in the latter the write is complete in quorum $Q_z$,
but the read has no way of knowing this.
We will use quorum views as a design element in our algorithms.

%% file: erato-SW.tex
\newcommand{\x}[1]{{\sc e}#1}

We now present and analyze the SWMR algorithm \algsw{}.

\vspace{-\medskipamount}
\subsection{Algorithm Description}
\label{sub:erato:sw:desc}

In \algsw{} reads take either \emph{two} or \emph{three} exchanges. 
%
%
%
This is achieved by combining the \emph{three} exchange read protocol 
of~\cite{HNS17} with the use of  Quorum Views of~\cite{GNS08}.
The read protocol design aims to return the value  associated with the timestamp 
of the last complete write operation.
We refer to the three exchanges of the read protocol as \x{1},
\x{2}, and \x{3}. Exchange \x{1} is initiated by the reader, and
exchanges \x{2} and \x{3} are conducted by the servers.
When the reader receive messages during \x{2}, it 
analyses the timestamps to  determine
whether to terminate or wait for the conclusion of \x{3}. 
Due to asynchrony it is possible for the message from \x{3}
to arrive at the reader before messages from \x{2}.
In this case the reader still terminates in \emph{three} exchanges.
%
%
Similarly to \ABD{}, writes take \emph{two} exchanges.
The code is given in Algorithm~\ref{alg:erato-sw}. We now give the details of the protocols; in referring to
the numbered lines of code we use the prefix ``L'' to stand for ``line''.


\begin{figure}[t!h]
\vspace{\bigskipamount}
\begin{minipage}{1.00\linewidth}
\begin{algorithm}[H] 
\caption{\small Reader, Writer, and Server Protocols for SWMR algorithm \algsw{}}
\label{alg:erato-sw}
\begin{multicols}{2}
{\sf\footnotesize%
\begin{algorithmic}[1]
\State {At each reader $r$}
\State \textbf{Variables:}
\State ${\it minTS}  , {\it maxTS}  \in \mathbb{N}; read\_op \in \mathbb{N}$ init 0
\State $RR  ,  RA  ,  maxACK \subseteq \srvSet \times M$
\State $v \in V$; $RRsrv  , RAsrv , maxTSrv \subseteq \srvSet$
\Function{Read}{}   
	\State $read\_op \gets read\_op + 1$ 		\label{line:erato:reader-bdcast:start}   
	\State $(RR , RA,RRsrv , RAsrv) \gets (\emptyset ,  \emptyset , \emptyset ,  \emptyset )$ 		   
	\State \textbf{bcast} $(\tup{{\sf readRequest}, r, read\_op})$ to $\srvSet$ 		
	\State \textbf{wait until} {$\exists Q \in \mathbb{Q} : (Q \subseteq RRsrv)  \lor (~Q \subseteq {\it RAsrv})$} \label{line:erato:reader-bdcast:end}    			          
	\If{$(\exists Q \in \mathbb{Q} : Q \subseteq RAsrv)$}		\label{line:erato:reader-3exch:start}	
		\State ${\it minTS} \gets $ \sf{min}$\{(m.ts) : (s,m) \in RA \land ~s \in Q \}$
		\State \textbf{return}($m.v$ s.t. $(s,m)\in RA \land ~ m.ts=minTS$)\label{line:erato:reader-3exch:end}	
	\ElsIf{$(\exists Q \in \mathbb{Q} : Q \subseteq RRsrv)$}
		\State $maxTS \gets $\sf{max}$(\{(m.ts) ~: $
		\State ~ ~ ~ ~ ~ ~ ~ ~ ~ ~ $(s,m) \in RR \land~s \in Q \})$	
		\State $maxACK \gets \{(s,m)  \in RR ~: $
		\State ~ ~ ~ ~ ~ ~ ~ ~ $s \in Q ~\land m.ts = maxTS \}$		
		\State $maxTSrv \gets \{s \in Q ~: (s,m) \in {\it maxACK} \}$
		\If{$(Q \subseteq maxTSrv$)}\label{line:erato:reader-qview1:start} 
			\State \textbackslash \textbackslash ** Qview1**\textbackslash \textbackslash
			\State \textbf{return}($m.v$ s.t. ({\it s,m}) $\in$ {\it maxACK})\label{line:erato:reader-qview1:end}
		\EndIf
		\If{$(\exists Q^{\prime} \in \mathbb{Q}, Q^{\prime} \ne Q : Q^{\prime} \cap Q \subseteq {\it maxTSrv})$}\label{line:erato:reader-qview3:start}
			\State \textbackslash \textbackslash ** Qview3**\textbackslash \textbackslash
			\State \textbf{wait until} {$\exists Q \in \mathbb{Q} : Q \subseteq RAsrv$}		
			\State $minTS \gets $ \sf{min}$(\{(m.ts)~ :$ 
			\State ~ ~ ~ ~ ~ ~ ~ ~ ~ ~ $(s,m) \in RA \land  ~s \in Q \})$
			\State \textbf{return$(m.v$ s.t. $(s,m) \in RA$
			\State ~ ~ ~ ~ ~ ~ $  \land ~s \in Q \land m.ts=minTS)$}  \label{line:erato:reader-qview3:end}
		\Else ~ // ** Qview2** //  \label{line:erato:reader-qview2:start}
			\State $\it{maxACK} \gets \{(s,m) \in RR ~: $
			\State ~ ~ ~ ~ ~ $ s \in Q \land ~ m.ts = {\it maxTS}-1 \}$
			\State \textbf{return$(m.v$ s.t. $(s,m) \in maxACK)$}		\label{line:erato:reader-qview2:end}
		\EndIf
	\EndIf
\EndFunction
\State \textbf{Upon receive} $m$ from $s$
  \If{$m.read\_op = read\_op$}
	 \If{$m.type = {\sf readRelay}$}
		\State $RR \gets RR \cup \{(s,m)\}$
		\State $RRsrv \gets RRsrv \cup \{s\}$
	\Else ~ // {\sf readAck} 	//
		\State $RA \gets RA \cup \{(s,m)\}$
		\State $RAsrv \gets RAsrv \cup \{s\}$
	\EndIf
\EndIf
\pagebreak
\State {At writer $w$}
\State \textbf{Variables:}
\State $ts \in \mathbb{N}^{+}$, $v\in V$, $wAck \subseteq \srvSet $
\State \textbf{Initialization:}
\State $ts \gets 0$, $v \gets \perp$, $wAck \gets \emptyset$
\Function{Write}{$val: input$}
	\State $(ts,v) \gets (ts+1, val)$   \label{line:erato:writer:start}    
	\State $wAck \gets \emptyset$ 
	\State \textbf{bcast} $(\tup{{\sf writeRequest}, ts, v, w})$ to $\srvSet$ 
	\State \textbf{wait until} {$(\exists Q \in \mathbb{Q} : Q \subseteq wAck)$}			
	\State \textbf{return}$()$ \label{line:erato:writer:end}    
\EndFunction
\State \textbf{Upon receive} $m$ from $s$
  \If{$m.ts = ts$}
	\State  $wAck \gets wAck \cup \{ s \}$
\EndIf
\smallskip
\State {At server $s$}
\State \textbf{Variables:}
\State $ts \in \mathbb{N}$ init 0 ;  $v \in V$ init $\bot$
\State $D \subseteq \srvSet$ init $\{s' : Q \in \mathbb{Q} \land (s, s' \in Q) \}$
\State $operations$ :  $\rdSet \rightarrow \mathbb{N}$ init $0^{|\rdSet|}$
\State $relays$ : $\rdSet \rightarrow  2^{\srvSet}$ init $0^{|\rdSet|}$
\smallskip
\State \textbf{Upon receive}{$(\tup{{\sf readRequest}, r, read\_op})$}{}     \label{line:erato:srv:readrequest:start}
	\State \textbf{bcast}$(\tup{{\sf readRelay}, ts, v, r, read\_op, s})$ to $D \cup r$ \label{line:erato:srv:readrequest:end}
\newline
\State \textbf{Upon receive}$(\tup{{\sf writeRequest}, ts^{\prime}, v^{\prime}, w})$ \label{line:erato:srv:write:start}
	\If{$(ts < ts^{\prime})$}																		\label{line:erato:srv:write-ts:start}
		\State $(ts,v)$ $\gets$ $(ts^{\prime}, v^{\prime})$								\label{line:erato:srv:write-ts:end}
	\EndIf													
	\State \textbf{send} $(\tup{writeAck,ts,s})$ to $w$ 								 \label{line:erato:srv:write:end}
\newline
\State \textbf{Upon receive}$(\tup{{\sf readRelay}, ts^{\prime}, v^{\prime}, r, read\_op, s})$
	~ \If{ $(ts < ts^{\prime})$}				\label{line:erato:srv:relay-ts:start}
		\State $(ts,v,vp)$ $\gets$ $(ts^{\prime},v^{\prime})$	\label{line:erato:srv:relay-ts:end}
	\EndIf								
	~ \If{ $(operations[r] < read\_op)$}  			  \label{line:erato:srv:relay-newop:start}
		\State $ operations[r] \gets read\_op $
		\State $ relays[r] \gets \emptyset $.          \label{line:erato:srv:relay-newop:end}
	\EndIf 								
	~ \If{$(operations[r] = read\_op)$} 		                \label{line:erato:srv:relay-readop:start}
		\State $relays[r] \gets relays[r] \cup \{s\}$   \label{line:erato:srv:relay-readop:end}
	\EndIf								
	~ \If{$(\exists Q \in \mathbb{Q} : Q \subseteq relays[r_i])$}					\label{line:erato:srv:relay-ack:start} 
		\State \textbf{send} $(\tup{{\sf readAck},ts, v, read\_op, s})$ to $r$	\label{line:erato:srv:relay-ack:end} \label{line:erato:srv:send:readack}
	\EndIf
\end{algorithmic}
}
\end{multicols}
\end{algorithm}
\end{minipage}
\vspace{\bigskipamount}
\end{figure}

\ppp{\em Reader Protocol.} 
Each reader $r$ maintain several temporary variables.
Key variable include $minTS$ and $maxTS$ hold the minimum and the maximum timestamps
discovered during the read operation.
Sets $RR$ and $RA$ hold the received {\sf readRelay} and {\sf readAck} messages respectively.
The ids of servers that sent these messages
are stored in  sets $RRsrv$ and $RAsrv$ respectively.
The set $maxTSrv$ keeps the ids of the servers that sent a {\sf readRelay} message with 
the timestamp equal to the maximum timestamp $maxTS$.

Reader $\rdr$ starts
its operation
by broadcasting a \textsf{readRequest} message
to the servers (exchange \x{1}). 
It then collects  \textsf{readRelay} messages
(from exchange \x{2}) and \textsf{readAck} messages (from exchange \x{3}).
The reader uses counter $read\_op$ to distinguish fresh message
from stale message from prior operations.
%
%
The  messages are collected until 
messages of the same type are received from some quorum $Q$ of servers
(L\ref{line:erato:reader-bdcast:start}-\ref{line:erato:reader-bdcast:end}).
%
%
If \textsf{readRelay} messages are received from  quorum $Q$ 
then the reader examines the timestamps
to determine what quorum view is observed (recall Section~\ref{sec:erato:model}).
%
If \vq{1} is observed,
then all timestamps are the same,
meaning that the write operation associated with the timestamp is complete,
 and it is safe 
	to return the value associated with it without exchange \x{3}.
	(L\ref{line:erato:reader-qview1:start}-\ref{line:erato:reader-qview1:end}).
%
If \vq{2} is observed, then the write associated with the maximum timestamp \emph{maxTS}
is not complete.
But because there is a sole writer,
it is safe to to return the value associated with timestamp \emph{maxTS-$1$}, i.e.,
the value of the preceding complete write, again without exchange \x{3}
	(L\ref{line:erato:reader-qview2:start}-\ref{line:erato:reader-qview2:end}).
%
If \vq{3} is observed, then the write associated with the maximum timestamp \emph{maxTS}
is in progress or complete. 
Since the reader is unable to decide which case applies,
it waits for the exchange \x{3} \textsf{readAck} messages from some quorum $Q$.
	%
The reader here returns the value associated with the \emph{minimum} timestamp observed
	(L\ref{line:erato:reader-qview3:start}-\ref{line:erato:reader-qview3:end}).
It is possible, due to asynchrony, that messages from \x{3} arrive from a quorum
before enough messages from \x{2} are gathered.
Here the reader decides as above for \x{3}
(L\ref{line:erato:reader-3exch:start}-\ref{line:erato:reader-3exch:end}).
%

\ppp{\em Writer Protocol.} 
Writer $w$ increments its local timestamp  $ts$  
and broadcasts a {\sf writeRequest} message 
to all  servers. 
It completes once  {\sf writeAck} messages are received from some quorum $Q$ 
(L\ref{line:erato:writer:start}-\ref{line:erato:writer:end}).

\ppp{\em Server Protocol.} 
Server $s$ stores the value of the replica $v$ and its associated timestamp $ts$.
The $relays$ array is used to store sets of processes that relayed to $s$
regarding a read operation. 
%
Destinations set $D$ is initialized to set containing all servers 
from every quorum that contains $s$. 
It is used for sending relay messages during exchange \x{2}.

In exchange \x{1} of a read, upon receiving message $\tup{{\sf readRequest}, r, read\_op}$, 
the server creates a {\sf readRelay} message, containing its $ts$, $v$, and $s$, 
and broadcasts it in exchange \x{2} to 
destinations in $D$ and  reader $r$
(L\ref{line:erato:srv:readrequest:start}-\ref{line:erato:srv:readrequest:end}).

In exchange \x{2},
upon receiving message $\tup{{\sf readRelay}, ts^{\prime}, v^{\prime}, r, read\_op}$ 
$\srvr$ compares its local timestamp $ts$ with $ts^{\prime}$.
If $ts < ts^{\prime}$, then $\srvr$ 
sets its local value and timestamp 
to those enclosed in the message
(L\ref{line:erato:srv:relay-ts:start}-\ref{line:erato:srv:relay-ts:end}).
Next, $\srvr$ checks if the received  {\sf readRelay} 
marks a new read by $r$,
i.e., $read\_op > operations[r]$.
If so, then $\srvr$: 
(a) sets its local counter for $r$ to the enclosed one, $operations[r] = read\_op$; 
and 
(b) re-initializes the relay set for $r$ to an empty set, $relays[r] = \emptyset$ 
(L\ref{line:erato:srv:relay-newop:start}-\ref{line:erato:srv:relay-newop:end}).
It then adds the sender of the  {\sf readRelay} message to the set of servers
that informed it regarding the read invoked by $r$
(L\ref{line:erato:srv:relay-readop:start}-\ref{line:erato:srv:relay-readop:end}).
Once  {\sf readRelay} messages are received from a  quorum $Q$,
$\srvr$ creates a 
 {\sf readAck} message
and sends it to $\rdr$ in exchange \x{3} of the read
(L\ref{line:erato:srv:relay-ack:start}-\ref{line:erato:srv:relay-ack:end}).

Within a write operation,  
upon receiving message $\tup{{\sf writeRequest}, ts', v', w}$, 
$\srvr$ compares its  $ts$ to the received one.
If $ts < ts'$, then $\srvr$ sets its local timestamp and value  to those
received, and sends acknowledgment to the  writer
(L\ref{line:erato:srv:write:start}-\ref{line:erato:srv:write:end}).

%
%
%
%

\vspace*{-\medskipamount}
\subsection{Correctness}
\label{sub:erato:sw:correct}

To prove correctness of algorithm \algsw{} 
we reason about its \emph{liveness} (termination) and \emph{atomicity} (safety).

\paragraph{\bf\em Liveness.} 
Termination is satisfied with respect to our failure model:
at least one quorum $Q$ is non-faulty 
and each operation waits for messages from a quorum $Q$ of servers.
Let us consider this in more detail.
\smallskip

\emph{Write Operation.} 
Showing \emph{liveness} is straightforward. 
Per algorithm \algsw{}, writer $w$ creates a \textsf{line:erato:writerequest} message 
and then it broadcasts it to all  servers.
Writer $w$ then waits for \textsf{writeAck} messages from a full quorum of servers 
(L\ref{line:erato:writer:start}-\ref{line:erato:writer:end}). 
Since in our failure model at least one quorum is non-faulty, then
writer $w$ collects \textsf{writeAck} messages form a full quorum of live servers 
and write operation $\omega$ terminates.
\smallskip

\emph{Read Operation.} 
The reader $\rdr$ begins by broadcasting a \textsf{readRequest} message all servers 
and waiting for responses. 
A read operation of the algorithm \algsw{} terminates when the reader $\rdr$ either
(i) collects \textsf{readAck} messages from full quorum of servers
or
(ii) collects \textsf{readRelay} messages from a full quorum and notices \vq{1} or \vq{2}
(L\ref{line:erato:reader-bdcast:start}-\ref{line:erato:reader-bdcast:end}).
Let's analyze case (i).
Since a full quorum $Q$ is non-faulty then 
at least a full quorum of servers receives the \textsf{readRequest} message. 
Any server $\srvr$ that receives this message broadcasts \textsf{readRelay} 
message to every server that belongs to the same quorum with, and the invoker $r_i$. 
That is its destinations set $D \cup \{r_i\}$ 
(L\ref{line:erato:srv:readrequest:start}-\ref{line:erato:srv:readrequest:end}).
In addition, no server ever discards any incoming \textsf{readRelay} messages. 
Any server, whether it is aware or not of the \textsf{readRequest}, 
always keeps a record of the incoming \textsf{readRelay} messages 
and takes action as if it is aware of the \textsf{readRequest}. 
The only difference between server $\srvr_i$ that received a \textsf{readRequest} message and server $\srvr_k$ that does not, is that $\srvr_i$ is able to broadcast \textsf{readRelay} messages, and $\srvr_k$  broadcasts \textsf{readRelay} messages when $\srvr_k$ receives the \textsf{readRequest} message. 
Each non-failed server receives \textsf{readRelay} messages from a full quorum of servers and 
sends a \textsf{readAck} message to reader $\rdr$ 
(L\ref{line:erato:srv:relay-ack:start}-\ref{line:erato:srv:relay-ack:end}).
Therefore, reader $\rdr$ can always collect \textsf{readAck} messages from a full quorum of servers, 
decide on a value to return, and terminate 
(L\ref{line:erato:reader-3exch:start}-\ref{line:erato:reader-3exch:end}). 
In case where case (ii) never holds then the algorithm will always terminate from case (i).
Thus, since any read or write operation will collect a {sufficient} number of messages and terminate 
then \emph{liveness} is satisfied.
 
Based on the above, it is always  the case that acknowledgment messages 
\textsf{readAck} and \textsf{writeAck} are collected from a full quorum of servers 
in any read
and write 
operation, thus ensuring  \emph{liveness}. 
 
\paragraph{\bf\em Atomicity.} 
To prove atomicity we order the operations 
with respect to the timestamps associated with the written values.
%
For each execution
of the algorithm there must
exist a partial order $\prec$ on the operations
that satisfy conditions A1, A2, and A3
given in Section~\ref{sec:erato:model}.
Let $ts_\pi$  
be the  the timestamp at the completion of $\pi$ 
when  $\pi$ is a write, 
and the timestamp associated with the returned value
when $\pi$ is a read.
%
We now define the partial order   as follows. 
For two operations $\pi_1$ and $\pi_2$, when $\pi_1$ is any operation and $\pi_2$ is a write, 
we let $\pi_1 \prec \pi_2$ if $ts_{\pi_1} < ts_{\pi_2}$.
For two operations $\pi_1$ and $\pi_2$,
when $\pi_1$ is a write and $\pi_2$ is a read we let $\pi_1 \prec \pi_2$ if $ts_{\pi_1} \leq ts_{\pi_2}$.
The rest of the order is established by transitivity,
without ordering the reads with the same timestamps.
We now state the following lemmas.

It is easy to see that the $ts$ variable in each server $s$ is monotonically increasing. 
This leads to the following lemma.


\begin{lemma}
\label{lem:erato:sw:srv:monotonic}
In any execution $\EX$ of \algsw{}, the variable $ts$ maintained by any server $\srvr$ in 
the system is non-negative and monotonically increasing.
\end{lemma}

\begin{Proof}
Upon receiving a timestamp $ts$, a server $\srvr$ updates its 
local timestamp $ts_s$ iff $ts > ts_s$, 
(L\ref{line:erato:srv:write-ts:start}-\ref{line:erato:srv:write-ts:end},\ref{line:erato:srv:relay-ts:start}-\ref{line:erato:srv:relay-ts:end}),
and the lemma follows.
\end{Proof}


Next, we show that any read operation that follows a write operation,
it receives \textsf{readAck} messages the servers where each included timestamp 
is at least as the one returned by the complete write operation.

\begin{lemma}
\label{lem:erato:sw:read-received-timestamps}
In any execution $\EX$ of \algsw{}, if a read operation $\rd$ succeeds a write 
operation $\omega$ that writes  $ts$ and $v$, i.e., $\omega \bef \rd$, and 
receives \textsf{readAck} messages from a quorum $Q$ of servers, set $RA$, then 
each $\srvr\in RA$ sends a \textsf{readAck} message to $\rd$ with a 
timestamp $ts_s$ s.t. $ts_s \geq ts$. 
\end{lemma}

\begin{Proof}
Let $wAck$ be the set of servers from a quorum $Q_a$ that send a \textsf{writeAck} message to $\omega$, 
let $RelaySet$ be the set of servers from a quorum $Q_b$ that sent \textsf{readRelay} messages to server $\srvr$,
and let $RA$ be the set of servers from a quorum $Q_c$ that send a \textsf{readAck} message to $\rd$. 
Notice that it is not necessary that $a \neq b \neq c$ holds.

Write operation $\omega$ is completed. 
By Lemma \ref{lem:erato:sw:srv:monotonic}, if a server $\srvr$ receives a timestamp $ts$ 
from a process $\pr$ at some time $T$, then $\srvr$ attaches a timestamp $ts^{\prime}$ 
s.t. $ts^{\prime} \geq ts$ in any message sent at any time $T^{\prime} \geq T$. 
Thus,
every server in $wAck$, sent a \textsf{writeAck} message to $\omega$ 
with a timestamp greater or equal to $ts$. 
Hence, every server $\srvr_x \in wAck$ has a timestamp $ts_{\srvr_x} \geq ts$. 
Let us now examine a timestamp $ts_s$ that server $\srvr$ sends to read operation $\rd$.  

Before server $\srvr$ sends a \textsf{readAck} message to $\rd$, 
it must receive \textsf{readRelay} messages from a full quorum $Q_b$ of servers, $RelaySet$
(L\ref{line:erato:srv:relay-ack:start}-\ref{line:erato:srv:relay-ack:end}).
Since both $wAck$ and $RelaySet$ contain messages from a full quorum of servers, 
and by definiton, any two quorums have a non-empty intersection, then $wAck\cap RelaySet\neq\emptyset$. 
By Lemma \ref{lem:erato:sw:srv:monotonic}, any server $\srvr_x \in aAck \cap RelaySet$ has a timestamp $ts_{s_x}$ s.t. $ts_{s_x} \geq ts$.
Since server $\srvr_x \in RelaySet$ and from the algorithm, 
server's $\srvr$ timestamp is always updated to the highest timestamp it noticed 
(L\ref{line:erato:srv:relay-ts:start}-\ref{line:erato:srv:relay-ts:start}),
then when server $\srvr$ receives the message from $\srvr_x$, 
it will update its timestamp $ts_s$ s.t. $ts_s \geq ts_{s_x}$.
Server $\srvr$ creates a \textsf{readAck} message where it encloses 
its local timestamp and its local value, $(ts_s, v_s)$
(L\ref{line:erato:srv:send:readack}).
Each $\srvr \in RA$ sends a \textsf{readAck} to $\rd$ with a timestamp $ts_s$ s.t. $ts_s \geq ts_{s_x} \geq ts$. 
Thus, $ts_s \geq ts$, and the lemma follows.
\end{Proof}


Now, we show that if a read operation succeeds a write operation, 
then it returns a value at least as recent as the one written.

\begin{lemma}
\label{lem:erato:sw:read-after-write}
In any execution $\EX$ of \algsw{}, if a read $\rd$ succeeds a write operation $\omega$ 
that writes timestamp $ts$, i.e. $\omega \to \rd$, and returns a timestamp $ts^{\prime}$, then $ts^{\prime} \geq ts$.      
\end{lemma}

\begin{Proof}
A read operation $\rd$ terminates when it either receives
(a) \textsf{readRelay} messages from a full quorum $Q$ 
or 
(b) \textsf{readAck} messages from a full quorum $Q$
(L\ref{line:erato:reader-bdcast:start}-\ref{line:erato:reader-bdcast:end}).

We first examine case (b). 
Let's suppose that $\rd$ receives \textsf{readAck} messages from a full quorum $Q$ of servers, $RA$. 
By lines 
\ref{line:erato:reader-3exch:start} - \ref{line:erato:reader-3exch:end}, 
it follows that $\rd$ decides on the minimum timestamp, $ts^{\prime}=minTS$, 
among all the timestamps in the \textsf{readAck} messages of the set $RA$.
From Lemma \ref{lem:erato:sw:read-received-timestamps}, $minTS \geq ts$ holds, 
where $ts$ is the timestamp written by the last complete write operation $\omega$. 
Then $ts^{\prime} = minTS \geq ts$ also holds. 
Thus, $ts^{\prime} \geq ts$. 

Now we examine case (a). 
In particular, case (a) terminates when the reader process notices either
(i) \vq{1} or (ii) \vq{2} or (iii) \vq{3}.
Let $wAck$ be the set of servers from a quorum $Q_a$ that send a \textsf{writeAck} message to $\omega$.
Since the write operation $\omega$, that wrote value $v$ associated with timestamp $ts$ is complete,
and by monotonicity of timestamps in servers (Lemma \ref{lem:erato:sw:srv:monotonic}),  
then at least a quorum $Q_a$ of servers has a timestamp $ts_a$ s.t. $ts_a \geq ts$.
In other words, 
every server in $wAck$, sent a \textsf{writeAck} message to $\omega$ 
with a timestamp $ts_a$ greater or equal to $ts$. 

Let's suppose that $\rd$ receives \textsf{readRelay} messages from a full quorum $Q_b$ of servers, $RR$. 
Since both $wAck$ and $RR$ contain messages from a full quorum of servers, quorums $Q_a$ and $Q_b$,
and by definition any two quorums have a non-empty intersection, then $wAck \cap RR \neq \emptyset$. 
Since every server in $wAck$ has a timestamp $ts_a \geq ts$ then 
any server  $s_x \in wAck \cap RR$ has a timestamp $ts_{s_x}$ s.t. $ts_{s_x} \geq ts_a \geq ts$.

If $\rd$ noticed \vq{1} in $RR$, 
then the distribution of the timestamps indicates the existence of one and only timestamp in $RR$, $ts^{\prime}$.
Hence, $ts^{\prime} \geq ts_{s_x} \geq ts_a \geq ts$.
Based on the algorithm 
(L\ref{line:erato:reader-qview1:start}-\ref{line:erato:reader-qview1:end}), 
the read operation $\rd$ returns value $v$ associated with $ts^{\prime}$
and $ts^{\prime} \geq ts$ holds.

Based on the definition of \vq{2}, if it is noticed in $RR$, 
then there must exist at least two servers in $wAck \cap RR$ with different timestamps 
and one of them holds the \emph{maximum} timestamp.
Let $s_k$ be the one that holds the \emph{maximum} timestamp $ts_{s_k}$ (or $maxTS$)
and $s_m$ the server that holds the timestamp $ts_{s_m}$ s.t. $maxTS = ts_{s_k} > ts_{s_m}$.
Since (a) any server $s_x \in wAck \cap RR$ has a timestamp $ts_{s_x}$ s.t. $ts_{s_x} \geq ts$, and
(b) $s_k \in wAck \cap RR$ holds the \emph{maximum} timestamp $ts_{s_k}$ (or $maxTS$), and
(c) $s_m \in wAck \cap RR$ and 
(d)  $maxTS = ts_{s_k} > ts_{s_m}$
then it follows that $maxTS = ts_{s_k} > ts_{s_m} \geq ts$. 
Thus, $ts_{s_k}$ (or $maxTS$) must be strictly greater from $ts$, $maxTS = ts_{s_k} > ts$.
Based on the algorithm, when $\rd$ notices \vq{2} in $RR$ 
then it returns the value $v$ associated with the \emph{previous maximum} timestamp,
that is the value associated with \emph{maxTS-1}
(L\ref{line:erato:reader-qview2:start}-\ref{line:erato:reader-qview2:end}).
Since $maxTS = ts_{s_k} > ts$, then for the \emph{previous maximum} timestamp, denoted by $ts^{\prime}$,
which is only one unit less than $maxTS$, then the following holds, $maxTS > maxTS-1 = ts^{\prime} \geq ts$.
Therefore, in this case $\rd$ returns a value $v$ associated with $ts^{\prime}$ and $ts^{\prime} \geq ts$ holds.

Lastly, when \vq{3} is noticed then $\rd$
waits for \textsf{readAck} messages from a full quorum $Q$ before termination, 	
(L\ref{line:erato:reader-qview3:start}-\ref{line:erato:reader-qview3:end}),
proceeds identically as in case (b) and the lemma follows.
\end{Proof}


\begin{lemma}
\label{lem:erato:sw:semifast-read-after-read}
In any execution $\EX$ of \algsw{}, if $\rd_1$ and $\rd_2$ 
are two \emph{semi-fast} read operations, take 3 exchanges to complete, 
such that $\rd_1$ precedes $\rd_2$, i.e., $\rd_1 \to \rd_2$, 
and $\rd_1$ returns the value for timestamp $ts_1$, then $\rd_2$ returns 
the value for timestamp $ts_2 \geq ts_1$.
\end{lemma}

\begin{Proof} 
Let the two operations $\rd_1$ and $\rd_2$ be invoked by processes with identifiers
$\rdr_1$ and $\rdr_2$ respectively (not necessarily different). Also, let $RA_1$ and 
$RA_2$ be the sets of servers from quorums $Q_a$ and $Q_b$ (not necessarily different)
that sent a \textsf{readAck} message to $\rdr_1$ and 
$\rdr_2$ during $\rd_1$ and $\rd_2$.

Assume by contradiction
that read operations $\rd_1$ and $\rd_2$ exist such that $\rd_2$ succeeds $\rd_1$, 
i.e., $\rd_1 \to \rd_2$, and the operation $\rd_2$ returns a timestamp $ts_2$ 
that is smaller than the $ts_1$ returned by $\rd_1$, i.e., $ts_2 < ts_1$. 
Based on the algorithm,  
$\rd_2$ returns a timestamp $ts_2$ that is smaller than 
the minimum timestamp received by $\rd_1$, i.e., $ts_1$, 
if $\rd_2$ obtains $ts_2$ and $v$ in the \textsf{readAck} 
message of some server $\srvr_x\in RA_2$, 
and $ts_2$ is the minimum timestamp received by $\rd_2$.

Let us examine if $\srvr_x$  
sends a \textsf{readAck} message to $\rd_1$ with timestamp $ts_x$,
i.e., $\srvr_x\in RA_1$. 
By Lemma \ref{lem:erato:sw:srv:monotonic}, and since $\rd_1\bef\rd_2$, then 
it must be the case that $ts_x \leq ts_2$.
According to our assumption $ts_1>ts_2$, and since $ts_1$ is the 
smallest timestamp sent to $\rd_1$ by any server in $RA_1$, then it follows that $\rdr_1$ does 
not receive the \textsf{readAck} message from $\srvr_x$, and hence $\srvr_x\notin RA_1$. 

Now let us examine the actions of the server $\srvr_x$. 
From the algorithm, server $\srvr_x$ collects \textsf{readRelay} messages from a full quorum $Q_c$
of servers before sending a \textsf{readAck} message to $\rd_2$
(L\ref{line:erato:srv:relay-ack:start}-\ref{line:erato:srv:relay-ack:start}).  
Let $RRSet_{\srvr_x}$ denote the set of servers from the full quorum $Q_c$ 
that sent \textsf{readRelay} to $\srvr_x$.
Since, both $RRSet_{\srvr_x}$ and $RA_1$ contain messages from full quorums, $Q_c$ and $Q_a$,
and since any two quorums have a non-empty intersection, 
then it follows that $RRSet_{\srvr_x}\cap RA_1\neq\emptyset$.

Thus there exists a server $\srvr_i \in RRSet_{\srvr_x}\cap RA_1$, 
that sent 
(i) a \textsf{readAck} to for $\rd_1$, 
and 
(ii) a \textsf{readRelay} to $\srvr_x$ during $\rd_2$. 
Note that $\srvr_i$ sends a \textsf{readRelay} for $\rd_2$ only after
it receives a read request from $\rd_2$. 
Since $\rd_1\bef \rd_2$, then it follows that $\srvr_i$ sent the 
\textsf{readAck} to $\rd_1$ before sending the \textsf{readRelay} to $\srvr_x$. 
By Lemma \ref{lem:erato:sw:srv:monotonic}, if $\srvr_i$ attaches a timestamp $ts_{s_i}$ 
in the \textsf{readAck} to $\rd_1$, then $\srvr_i$ attaches a timestamp $ts_{s_i}^{\prime}$ 
in the \textsf{readRelay} message to $\srvr_x$, such that  $ts_{s_i}^{\prime} \geq ts_{s_i}$ 
Since $ts_1$ is the minimum timestamp received by $\rd_1$, 
then $ts_{s_i} \geq ts_1$, and hence $ts_{s_i}^{\prime} \geq ts_1$ as well. 
By Lemma \ref{lem:erato:sw:srv:monotonic}, and since $\srvr_x$ receives the \textsf{readRelay} 
message from $\srvr_i$ before sending a \textsf{readAck} to $\rd_2$, 
it follows that $\srvr_x$ sends a timestamp $ts_2$ s.t. $ts_2 \geq ts_{s_i}^{\prime} \geq ts_1$.
Thus, $ts_2 \geq ts_1$ and this contradicts our initial assumption. 
\end{Proof}


\begin{lemma}
\label{lem:erato:sw:fast-read-after-read}
In any execution $\EX$ of \algsw{}, if $\rd_1$ and $\rd_2$ 
are two \textsf{fast} read operations, take 2 exchanges to complete, 
such that $\rd_1$ precedes $\rd_2$, i.e., $\rd_1 \to \rd_2$, 
and $\rd_1$ returns the value for timestamp $ts_1$, then $\rd_2$ returns 
the value for timestamp $ts_2 \geq ts_1$.
\end{lemma}

\begin{Proof} 
Let the two operations $\rd_1$ and $\rd_2$ be invoked by processes with identifiers
$\rdr_1$ and $\rdr_2$ respectively (not necessarily different). Also, let $RR_1$ and 
$RR_2$ be the sets of servers from quorums $Q_a$ and $Q_b$ (not necessarily different)
that sent a \textsf{readRelay} message to $\rdr_1$ and 
$\rdr_2$ during $\rd_1$ and $\rd_2$.

The algorithm terminates in \emph{two} communication exchanges
when a read operation $\rd$ receives \textsf{readRelay} messages 
from a full quorum $Q$ and based on the distribution of the timestamp
it either notices 
(a) \vq{1}
or 
(b) \vq{2}.
We now examine the four cases. 

Case (i), $\rd_1 \to \rd_2$ and both $\rd_1$ and $\rd_2$ notice \vq{1}. 
It is known that all the servers in $RR_1$ replied to $\rd_1$ with timestamp $ts_1$.
Since by definition, any two quorums have a non-empty intersection
it follows that $RR_1 \cap RR_2 \neq \emptyset$.
From that and by Lemma \ref{lem:erato:sw:srv:monotonic}, then every server
$s_x \in RR_1 \cap RR_2$ has a timestamp $ts^{\prime}$ such that
$ts^{\prime} \geq ts_1$.
Since $\rd_2$ notices \vq{1} in $RR_2$, 
then the distribution of the timestamps indicates the existence of one and only timestamp
in $RR_2$, $ts_2$. Thus, $ts_2 \geq ts^{\prime} \geq ts_1$.

Case (ii), $\rd_1 \to \rd_2$ and $\rd_1$ notices \vq{1} and $\rd_2$ notices \vq{2}. 
It is known that all the servers in $RR_1$ replied to $\rd_1$ with timestamp $ts_1$.
Since by definition, any two quorums have a non-empty intersection
it follows that $RR_1 \cap RR_2 \neq \emptyset$.
From that and by Lemma \ref{lem:erato:sw:srv:monotonic}, then every server
$s_x \in RR_1 \cap RR_2$ has a timestamp $ts^{\prime}$ such that
$ts^{\prime} \geq ts_1$.
Since $\rd_2$ notices \vq{2} in $RR_2$, 
then there must exist at least two servers in $RR_1 \cap RR_2$ 
with different timestamps and one of them holds the \emph{maximum} timestamp.
Let $s_k$ be the one that holds the \emph{maximum} timestamp $ts_{s_k}$ (or $maxTS$)
and $s_m$ the server that holds the timestamp $ts_{s_m}$ s.t. $maxTS = ts_{s_k} > ts_{s_m}$.
Since (a) any server $s_x \in RR_1 \cap RR_2$ has a timestamp $ts^{\prime}$ s.t. $ts^{\prime} \geq ts_1$, 
and
(b) $s_k \in RR_1 \cap RR_2$ holds the \emph{maximum} timestamp $ts_{s_k}$ (or $maxTS$), and
(c) $s_m \in RR_1 \cap RR_2$ and 
(d)  $maxTS = ts_{s_k} > ts_{s_m}$
then it follows that $maxTS = ts_{s_k} > ts_{s_m} \geq ts_1$. 
Thus, $ts_{s_k}$ (or $maxTS$) must be strictly greater from $ts_1$, $maxTS = ts_{s_k} > ts_1$.
Based on the algorithm, when $\rd$ notices \vq{2} in $RR_2$ 
then it returns the value $v$ associated with the \emph{previous maximum} timestamp,
that is the value associated with \emph{maxTS-1}
(L\ref{line:erato:reader-qview2:start}-\ref{line:erato:reader-qview2:end}).
Since $maxTS = ts_{s_k} > ts_1$, then for the \emph{previous maximum} timestamp, 
denoted by $ts_2$,
which is only one unit less than $maxTS$, 
then the following holds, $maxTS > maxTS-1 = ts_2 \geq ts_1$, thus $ts_2 \geq ts_1$.

Case (ii), $\rd_1 \to \rd_2$ and $\rd_1$ notices \vq{2} and $\rd_2$ notices \vq{1}.
Since $\rd_1$ notices \vq{2} in $RR_1$
then there exist a subset of servers $Smax$,  $Smax \subset RR_1$,
that hold the \emph{maximum} timestamp $maxTS$
and a subset of servers $Spre$,  $Spre \subset RR_1$,
that hold timestamp \emph{maxTS-1}. 
Based on the algorithm, $\rd_1$ returns $ts_1$ s.t. $ts_1 = maxTS-1$
from the set of servers in $Spre$.
Since $RR_1 \cap RR_2 \neq \emptyset$, and \vq{1} indicates
the existence of one and only timestamp, then $\rd_2$ can notice 
\vq{1} in two cases; 
(a) all the servers in $RR_1 \cap RR_2 \subseteq Spre$
or
(b) all the servers in $RR_1 \cap RR_2 \subseteq Smax$.
By Lemma \ref{lem:erato:sw:srv:monotonic}, and if (a) holds then $\rd_2$ returns $ts_2$ s.t. $ts_2 \geq ts_1$;
else, if (b) holds then $\rd_2$ returns $ts_2$ s.t. $ts_2 > ts_1$.

Case (i), $\rd_1 \to \rd_2$ and both $\rd_1$ and $\rd_2$ notice \vq{2}. 
The distribution of the timestamps that $\rd_1$ notices, 
indicates that the write operation associated with the \emph{maximum} timestamp, $maxTS$, 
is on-going, i.e., not completed.
By the property of \emph{well formdness} and the existence of a sole writer in
the system then we know that $ts_1$ corresponds to the latest complete write operation, $ts_1 = maxTS-1$.
By Lemma \ref{lem:erato:sw:read-after-write}, $\rd_2$ will not be able to return a timestamp $ts_2$
s.t. $ts_2 < maxTS-1$.
Thus $ts_2 \geq ts_1$ holds and the lemma follows.
\end{Proof}


\begin{lemma}
\label{lem:erato:sw:read-after-read}
In any execution $\EX$ of \algsw{}, if $\rd_1$ and $\rd_2$ 
are two read operations
such that $\rd_1$ precedes $\rd_2$, i.e., $\rd_1 \to \rd_2$, 
and $\rd_1$ returns 
timestamp $ts_1$, then $\rd_2$ returns 
a timestamp $ts_2$, s.t. $ts_2 \geq ts_1$.
\end{lemma}

\begin{Proof} 
We are interested to examine the cases where one of the read operation is 
\emph{fast} and the other is \emph{semifast}.
In particular, cases
(i) $\rd_1 \to \rd_2$ and $\rd_1$ is \emph{semifast} and $\rd_2$ is \emph{fast}
and
(ii) $\rd_1 \to \rd_2$ and $\rd_1$ is \emph{fast} and $\rd_2$ is \emph{semifast}.

Let the two operations $\rd_1$ and $\rd_2$ be invoked by processes with identifiers
$\rdr_1$ and $\rdr_2$ respectively (not necessarily different). 
Also, let $RR_1$,$RA_1$ and $RR_2$,$RA_2$ be the sets of servers 
from full quorums (not necessarily different)
that sent a \textsf{readRelay} and \textsf{readAck} message to $\rd_1$ and $\rd_2$ respectively.

We start with case (i). 
Since read operation $\rd_1$ is \emph{semifast}, then based on the algorithm,
the timestamp $ts_1$ that is returned it is also the \emph{minimum} timestamp noticed
in $RA_1$.
Before a server $s$ sents \textsf{readAck} messages to $\rd_1$ (that form $RA_1$),
it must receive \textsf{readRelay} messages from a full quorum of servers.
Thus, by Lemma \ref{lem:erato:sw:srv:monotonic} monotonicity of the timestamps at the servers
we know that the \emph{minimum} timestamp that a full quorum has by the end
of $\rd_1$ is $ts_1$.
Read operation $\rd_2$ receives \textsf{readRelay} messages from a full quorum of servers, $RR_2$.
By definition of quorums, since both $RA_1$ and $RR_2$ are from a full quorum of
servers then it follows that $RA_1 \cap RR_2 \neq \emptyset$.
Thus every server $s_x \in RA_1 \cap RR_2$ holds a timestamp $ts^{\prime}$ s.t. $ts^{\prime} \geq ts_1$.

If $\rd_2$ notices \vq{1} in $RR_2$ then the distribution of the timestamps
in $RR_2$ indicates the existance of one and only timestamp, $maxTS$.
From the above, it follows that for the timestamp $ts_2$ that $\rd_2$ returns 
$maxTS = ts_2 \geq ts^{\prime} \geq ts_1$ holds. 

On the other hand, if $\rd_2$ notices \vq{2} in $RR_2$, 
then based on the distributions of the timestamps  in \vq{2} 
there must exist at least two servers in $RA_1 \cap RR_2$ 
with different timestamps and the one must be the \emph{maximum}.
Since every server $s_x \in RA_1 \cap RR_2$ holds a timestamp $ts^{\prime}$ s.t. $ts^{\prime} \geq ts_1$
then the \emph{maximum} timestamp $maxTS$ cannot be 
equal to $ts_1$. If that was the case, $\rd_2$ would have noticed \vq{1}.
In particular, now $maxTS > ts_1$ holds.
Based on the algorithm, when $\rd$ notices \vq{2} in $RR_2$ 
then it returns the value $v$ associated with the \emph{previous maximum} timestamp,
that is the value associated with \emph{maxTS-1}
(L\ref{line:erato:reader-qview2:start}-\ref{line:erato:reader-qview2:end}).
Since $maxTS  > ts_1$, then for the \emph{previous maximum} timestamp, 
denoted by $ts_2$,
which is only one unit less than $maxTS$, 
then the following holds, $maxTS > maxTS-1 = ts_2 \geq ts_1$, thus $ts_2 \geq ts_1$.

We now examine
case (ii). 
Since $\rd_1$ is fast, it follows that it has either noticed \vq{1} or \vq{2} in $RR_1$. 
If \vq{1} was noticed, 
and $\rd_1$ returned a value associated with \emph{maximum} timestamp $ts_1$, 
then by the completion of $\rd_1$
a full quorum has a timestamp $ts^{\prime}$ s.t. $ts^{\prime} \geq ts_1$.
Now, since read operation $\rd_2$ is \emph{semifast}, then based on the algorithm,
the timestamp $ts_2$ that is returned it is the \emph{minimum} timestamp noticed
in $RA_2$.
Before a server $s$ sents \textsf{readAck} messages to $\rd_2$ (that form $RA_2$),
it must receive \textsf{readRelay} messages from a full quorum of servers, $RelaySet$.
By Lemma \ref{lem:erato:sw:srv:monotonic} monotonicity of the timestamps at the servers
and $RR_1 \cap RelaySet \neq \emptyset$, then every server in $RA_2$
has a timestamp $ts_2$ s.t. $ts_2 \geq ts^{\prime} \geq ts_1$.

If \vq{2} was noticed in $RR_1$, 
based on the algorithm, $\rd_1$ returned a value associated with 
\emph{previous maximum} timestamp, that is $ts_1$. 
By the completion of $\rd_1$ a full quorum has a timestamp 
$ts^{\prime}$ s.t. $ts^{\prime} \geq ts_1$.
Read operation $\rd_2$ is \emph{semifast}, 
and the returned timestamp $ts_2$ is the \emph{minimum} 
timestamp noticed in $RA_2$.
A server $s$ sents \textsf{readAck} messages to $\rd_2$ (that form $RA_2$),
when receives \textsf{readRelay} messages from a full quorum of servers, $RelaySet$.
By Lemma \ref{lem:erato:sw:srv:monotonic} 
and since $RR_1 \cap RelaySet \neq \emptyset$, 
then every server in $RA_2$ has a timestamp $ts_2$ s.t. $ts_2 \geq ts^{\prime} \geq ts_1$.

The rest of the cases are proved in Lemmas \ref{lem:erato:sw:semifast-read-after-read} and \ref{lem:erato:sw:fast-read-after-read}
and the lemma follows.
\end{Proof}

\begin{theorem}
Algorithm {\small \sc \algsw{}} implements an atomic {\small SWMR} object.
\end{theorem}
\begin{Proof} 
We now use the lemmas  above and the partial order definition
to reason about each of the three conditions A1, A2 and A3. 

\ppp{A1}  For any $\op_1,\op_2\in\Pi$ such that $\op_1\bef\op_2$, it cannot be that $\op_2\prec\op_1$.
		
\noindent 	When the two operations $\op_1$ and $\op_2$ are reads and $\op_1\bef\op_2$ holds, then	
				from Lemma~
				\ref{lem:erato:sw:read-after-read}
				it follows that the timestamp of $\op_2$ is no less than the one rof $\op_1$,
				$ts_{\op_2} \geq ts_{\op_1}$.
				If $ts_{\op_2} > ts_{\op_1}$ then by the ordering definition $\op_1\prec\op_2$ is satisfied.
				When $ts_{\op_2} = ts_{\op_1}$ then the ordering is not defined, thus it cannot be the case that 
				$\op_2\prec\op_1$.
				If $\op_2$ is a write, the sole writer generates a new  timestamp by
				incrementing the largest timestamp in the system. 
				By well-formedness (see Section \ref{sec:erato:model}), 
					any timestamp generated
					 in any write operation that 
				precedes $\op_2$ must be smaller than $ts_{\op_2}$.
				Since $\op_1\bef\op_2$, then it holds that $ts_{\op_1} < ts_{\op_2} $. Hence, by the ordering
				definition it cannot be the case that $\op_2\prec\op_1$.
				Lastly, when $\op_2$ is a read and $\op_1$ a write and $\op_1\bef\op_2$ holds, 
				then from Lemma~
				\ref{lem:erato:sw:read-after-write}
				it follows that 
				$ts_{\op_2} \geq ts_{\op_1}$.
				By the ordering definition, it cannot hold that $\op_2\prec\op_1$ in this case either.

\ppp{A2} For any write $\wrt\in\Pi$ and any operation $\op\in\Pi$, then either $\wrt\prec\op$ or $\op\prec\wrt$.

\noindent	If the timestamp returned from $\wrt$ is greater than the one returned from $\op$,
				i.e. $ts_{\wrt} > ts_{\op}$,
				 then $\op\prec\wrt$ follows directly.
				Similartly, if $ts_{\wrt} < ts_{\pi}$ holds, then $\wrt\prec\op$ follows.
				If $ts_{\wrt} = ts_{\pi}$, then it must be that $\op$ is a read 
				and 
				$\op$ discovered $ts_{\wrt}$ in a quorum view \vq{1} or \vq{3}.	
				Thus, $\wrt\prec\op$ follows.
			
\ppp{A3} Every read operation returns the value of the last write preceding it according to $\prec$ (or the initial value if there is no such write).
	
\noindent	Let $\wrt$ be the last write preceding read $\rho$.
				From our definition it follows that $ts_{\rho} \geq ts_{\wrt}$.
		    		If $ts_{\rho} = ts_{\wrt}$, then $\rho$ returns
		   		 the value conveyed by $\wrt$ to some servers in a quorum $Q$,
		   		 satisfying either $\vq{1}$ or $\vq{3}$. 
		   		 %
		    		%
		    		If $ts_{\rho} > ts_{\wrt}$, then $\rho$ obtains a larger 
		    		timestamp, but such a timestamp can only be created by a write
		    		that succeeds $\wrt$, thus $\wrt$ does not precede the read and this cannot be the case.
		    		Lastly, if $ts_{\rho} = 0$, no preceding writes exist, and $\rho$ returns the initial value.
\end{Proof}

Having shown liveness and atomicity of algorithm \algsw{} the result follows.

\subsection{Performance}
\label{sub:erato:sw:performance}

We now assess the performance of \algsw{} in terms of 
(i) latency of read and write operations as measured by the number of communication exchanges, and 
(ii) the message complexity of read and write operations.

\ppp{Communication and Message Complexity.} 
By inspection of the code, write operations take {2} \metrics{} and
read operations take either {2} or {3} \metrics{}.
The (worst case) message complexity of 
write operations is $2|\srvSet|$
and of read operations
is $|\srvSet|^{2} + 2|\srvSet|$,
as follows from the structure of the algorithm.
We now give additional details.

\ppp{Operation Latency.}
\emph{Write operation latency:} 
According to algorithm \algsw{}, writer $w$ sends \textsf{line:erato:writerequest} messages 
to all servers during exchange \x{1} 
and waits for \textsf{writeAck} messages from a full quorum of servers during \x{2}.
Once the \textsf{writeAck} messages are received, no further communication is required and the write operation
terminates. Therefore, any write operation  consists of \emph{2} communication exchanges. 

\emph{Read operation latency:} 
A reader sends a \textsf{readRequest} message to all the servers
in the first communication exchange \x{1}. 
Once the servers receive the \textsf{readRequest}
message they broadcast a \textsf{readRelay} message to 
all servers \emph{and} the reader in exchange \x{2}.
The reader can terminate at the end of the \x{2}
if it receives \textsf{readRelay} messages and based on the distribution of 
the timestamp it notices \vq{1} or \vq{2} 
%
If this is not the case, the operation goes into the third exchange \x{3}.
Thus read operations terminate after either \emph{2} or \emph{3} communication exchanges.

\ppp{Message Complexity.}
We measure operation message complexity as the worst case number of exchanged messages 
in each read and write operation. 
The worst case number of messages corresponds to failure-free executions where all
participants send messages to all destinations according to the protocols. 

\emph{Write operation:} 
A single write operation in algorithm \algsw{} takes \emph{2} communication exchanges.
In the first exchange \x{1}, the writer sends a {\sf{writeRequest}} message to all the servers in $\srvSet$.
The second exchange \x{2}, occurs when all servers in $\srvSet$ send a {\sf{writeAck}} message 
to the writer. 
Thus, at most $2|\srvSet|$ messages are exchanged in a write operation.

\emph{Read operation:} 
Read operations in the worst case take \emph{3} communication exchanges. 
Exchange \x{1} occurs when a reader sends a {\sf{readRequest}} message 
to all servers in $\srvSet$. 
The second exchange \x{2} occurs when servers in $\srvSet$ send {\sf{readRelay}}
messages to all servers in $\srvSet$ and to the requesting reader. 
The final exchange \x{3} occurs when servers in $\srvSet$ send 
a {\sf{readAck}} message to the reader. 
Summing together $|\srvSet| + (|\srvSet|^{2} + |\srvSet|)+ |\srvSet|$, 
shows that in the worst case, $|\srvSet|^{2} + 3|\srvSet|$ 
messages are exchanged during a read operation.

%% file: erato-MW.tex
We now aim for a MWMR algorithm that involves \emph{two} or \emph{three} communications exchanges per read operation
and \emph{four} exchanges per write operation.
%
The read protocol 
of algorithm \algsw{} 
relies on the fact of the sole writer in the system: 
based on the distribution of the timestamp in a quorum $Q$,
if the reader knows that the write operation is  not complete, 
then any previous write is complete (by well-formedness).
In the MWMR setting this does not hold  due to the possibility of concurrent writes.
%
%
Consequently, algorithm \algmw{}, in order to allow
operations to terminate in either \emph{two} or \emph{three} communication exchanges,
adapts the read protocol from algorithm \omam{} in combination with the
iterative technique using quorum views of \cwfr{}.
The latter approach not only predicts 
the completion status of a write operation, but also detects the last potentially complete write operation. 
The code is given in in Algorithm~\ref{alg:erato-mw}.


\begin{figure}[t!h]
\vspace{-\bigskipamount}
\begin{minipage}{1\linewidth}
\begin{algorithm}[H]
\caption{\small Reader, Writer and Server Protocols for MWMR algorithm \algmw{}}
\label{alg:erato-mw}
\begin{multicols}{2}
{\sf\footnotesize
\begin{algorithmic}[1]
\State {At each reader $r$}
\State \textbf{Variables:}
\State $v \in V$; $read\_op \in \mathbb{N}$; $minTAG , maxTAG \in T$ 
\State $RR , RA, maxACK \subseteq \srvSet \times M$ init $\emptyset$
\State  ${\it RRsrv , RAsrv , maxTGsrv} \subseteq \srvSet$ init $\emptyset$
\State \textbf{Initialization:}
\State $minTAG \gets \tup{0,0}$, $maxTAG \gets \tup{0,0}$
\State $v \gets \perp, read\_op \gets 0$
\Function{Read}{}
	\State $read\_op \gets read\_op + 1$ 					          \label{line:erato:mw:bdcast:start}
	\State $(RR , RA, maxACK) \gets (\emptyset ,  \emptyset , \emptyset )$ 		
	\State $(RRsrv , RAsrv, maxTGsrv) \gets (\emptyset ,  \emptyset , \emptyset)$ 		
	\State \textbf{bcast} $(\tup{{\sf readRequest}, r, read\_op})$ to $\srvSet$ 
	\State \textbf{wait until}{$(\exists Q \in \mathbb{Q} : Q \subseteq RRsrv \lor Q \subseteq RAsrv)$}		\label{line:erato:mw:bdcast:end}
	\If{$(\exists Q \in \mathbb{Q} : Q \subseteq RAsrv)$}			\label{line:erato:mw:3exch:start}
		\State $minTAG \gets$ \sf{min}$(\{(m.ts,m.id):$ 
		\State ~ ~ ~ ~ ~ ~ ~ ~ ~ ~ ~ ~ $(s,m) \in RA ~\land s \in Q\})$
		\State \textbf{return}$(m.v$ s.t. $(s,m) \in RA \land s \in Q $
		\State ~ ~ ~ ~ ~ ~ $\land~ (m.ts, m.id)=minTAG)$									\label{line:erato:mw:3exch:end}
	\ElsIf{$(\exists Q \in \mathbb{Q} : Q \subseteq RRsrv)$} 
		\While{$(Q \neq \emptyset)$}   \label{line:erato:mw:reader:unravel:start}
			\State ${\it maxTAG} \gets$ \sf{max}$\{(m.ts,m.id):$ 
			\State ~ ~ ~ ~ ~ ~ ~ ~ ~ ~ $(s,m) \in RA ~\land s \in Q\}$				
			\State ${\it maxACK} \gets \{(s,m) \in RR : s \in Q \land $
			\State ~ ~ ~ ~   ${\it (m.ts, m.id) =  maxTAG}\}$	
			\State ${\it maxTGsrv} \gets \{s \in Q$ :
			\State ~ ~ ~ ~ ~ ~ ~ ~ ~ ~  $(s,m) \in maxACK \}$	
			\If{$Q \subseteq maxTGsrv$}  \label{line:erato:mw:reader:qview1:start}
				\State //** Qview1**//
				\State \textbf{return}$(m.v$ s.t. $(s,m) \in maxACK)$
			\EndIf											\label{line:erato:mw:reader:qview1:end}
			\If{$(\exists Q' \in \mathbb{Q}, Q' \neq Q : Q' \cap Q \subseteq {\it maxTGsr})$} \label{line:erato:mw:reader:qview3:start}
				\State // ** Qview3**//
				\State \textbf{wait until ~}{$ (\exists Q^{\prime\prime} \in \mathbb{Q} : Q^{\prime\prime} \subseteq RAsrv)$}		
				\State ${\it minTAG} \gets$ \sf{min}$\{(m.ts,m.id):$
				\State ~ ~ ~ ~ ~ ~ ~ ~  ${\it (s,m) \in RA ~\land s \in Q^{\prime\prime}}\}$
				\State \textbf{return}$(m.v$ s.t., $(s,m) \in RA \land s \in Q''$
				\State ~ ~ ~ ~ ~ $\land ~{\it (m.ts, m.id)=minTAG})$  \label{line:erato:mw:reader:qview3:end}
			\Else ~//** Qview2**// \label{line:erato:mw:reader:qview2:start}
				\State $Q \gets Q \setminus maxTGsrv$	
			\EndIf \label{line:erato:mw:reader:qview2:end} \label{line:erato:mw:reader:unravel:end}
		\EndWhile
	\EndIf
\EndFunction
\newline
\State \textbf{Upon receive} $m$ from $s$
  \If{$m.read\_op = read\_op$}
	 \If{$m.type = {\sf readRelay}$}
		\State $RR \gets RR \cup \{(s,m)\}$
		\State $RRsrv \gets RRsrv \cup \{s\}$
	\Else ~ // {\sf readAck} 	//
		\State $RA \gets RA \cup \{(s,m)\}$
		\State $RAsrv \gets RAsrv \cup \{s\}$
	\EndIf
\EndIf
\smallskip
\pagebreak
\State {At each writer $w$}
\State \textbf{Variables:}
\State $ts \in \mathbb{N}$, $v \in V$, $write\_op \in \mathbb{N}$, $maxTS \in \mathbb{N}$ 
\State $Acks \subseteq \srvSet \times M$ init $\emptyset$ ; $AcksSrv \subseteq \srvSet$ init $\emptyset$
\State \textbf{Initialization:}
\State $ts \gets 0$, $v \gets \perp$, $write\_op \gets 0$, $maxTS \gets 0$
\Function{Write}{$val: input$}
	\State $write\_op \gets write\_op +1$ \label{line:erato:mw:writer:discover:start}
	\State $(Acks, AcksSrv) \gets (\emptyset,\emptyset) $
	\State \textbf{bcast} $(\tup{ {\sf writeDiscover}, write\_op, w})$ to $\srvSet$ 
	\State \textbf{wait until} {$(\exists Q \in \mathbb{Q} : Q \subseteq AcksSrv)$}  \label{line:erato:mw:writer:discover:end}
	\State ${\it maxTS} \gets$ \sf{max}$\{(m.ts) :$
	\State ~ ~ ~ ~ ~ ~ ~ $ (s,m) \in Acks \land s \in Q \}$	\label{line:erato:mw:writer:ph2:start} \label{line:erato:mw:writer:find-max}
	\State $(ts,id,v) \gets (maxTS+1, i ,val)$ \label{line:erato:mw:writer:create-tag}
	\State $write\_op \gets write\_op +1$
	\State $(Acks, AcksSrv) \gets (\emptyset,\emptyset) $
	\State \textbf{bcast} $(\tup{{\sf writeRequest}, ts, v, w, write\_op})$ to $\srvSet$ 	 
	\State \textbf{wait until} {$(\exists Q \in \mathbb{Q} : Q \subseteq AcksSrv)$}      \label{line:erato:mw:writer:ph2:end}			
	\State \textbf{return}$()$
\EndFunction
\State \textbf{Upon receive} $m$ from $s$
  \If{$m.write\_op = write\_op$}
		\State $Acks \gets Acks \cup \{(s,m)\}$
		\State $AcksSrv \gets AcksSrv \cup \{s\}$
\EndIf
\smallskip
\State {At server $s$}
\State \textbf{Variables and Initialization:}
\State $ts \in \mathbb{N}$ init 0; $id \in \wSet$ init $\bot$; $v \in V$ init $\bot$
\State ${\it operations}$ : $\rdSet \rightarrow \mathbb{N}$ init $0^{|\rdSet|}$
\State  ${\it write\_ops}$ :  $\wSet \rightarrow \mathbb{N}$  init $0^{|\wSet|}$
\State $relays$ :  $\rdSet \rightarrow 2^{\srvSet}$  init $\emptyset^{|\rdSet|}$
\State $D \subseteq \srvSet$ init $\{s : (\exists Q \in \mathbb{Q}), (s, s_i\in Q) \}$ 
\smallskip
\State \textbf{Upon receive}$(\tup{{\sf writeDiscover }, write\_op, w})$		\label{line:erato:mw:srv:discover:start}											
	\State \textbf{send} $(\tup{{\sf discoverAck}, ts, id, s_i})$ to $w$ \label{line:erato:mw:srv:discover:end}					
\smallskip
\State \textbf{Upon receive}
\State ~ ~{$(\tup{{\it {\sf writeRequest}, {\it ts', v', id', write\_op,w}}})$} \label{line:erato:mw:srv:write-p2:start}													
	\If{$write\_ops[w] < write\_op$}																		
		\State $write\_ops[w] \gets write\_op$
		\If{$(ts < ts^{\prime}) \lor (ts = ts^{\prime} \land id < id^{\prime})$ }				\label{line:erato:mw:srv:relay-tg:start}
			\State $(ts,id,v)$ $\gets$ $(ts^{\prime}, id^{\prime}, v^{\prime})$				\label{line:erato:mw:srv:relay-tg:end}
		\EndIf		
	\EndIf											
	\State \textbf{send} $(\tup{{\sf writeAck}, write\_op, s})$ to $w$ 	\label{line:erato:mw:srv:write-p2:end}
\smallskip
\State \textbf{Upon receive}{$(\tup{{\sf readRequest}, r, read\_op})$}{} \label{line:erato:mw:srv:relay-ack:start}
	\State \textbf{bcast}$ \tup{{\it {\sf readRelay}, ts, id, v, r, read\_op, s}}$ to $D \cup r$ \label{line:erato:mw:srv:relay-ack:end}
\smallskip
\State \textbf{Upon receive}$(\langle${\sf readRelay},{\it ts',id',v',r,read\_op,s}$\rangle)$
	~ \If{$(ts < ts^{\prime}) \lor (ts = ts^{\prime} \land id < id^{\prime})$ }		\label{line:erato:mw:srv:relay-tag:start}
			\State $(ts,id, v)$ $\gets$ $(ts^{\prime},id^{\prime},v^{\prime})$			\label{line:erato:mw:srv:relay-tag:end}
	\EndIf								
	~ \If{ $(operations[r] < read\_op)$}  			
		\State $ operations[r_i] \gets read\_op $ ;  $ relays[r] \gets \emptyset $.
	\EndIf 								
	~ \If{$(operations[r] = read\_op)$} 		
		\State $relays[r] \gets relays[r] \cup \{s\}$ 
	\EndIf								
	~ \If{$(\exists Q \in \mathbb{Q} : Q \subseteq relays[r])$}		\label{line:erato:mw:srv:read-ack:start}
		\State \textbf{send} $(\tup{{\sf readAck},ts, id, v, read\_op, s})$ to $r$ \label{line:erato:mw:srv:read-ack:end}\label{line:erato:mw:srv:send:readack}
	\EndIf										
\end{algorithmic}
}
\end{multicols}\vspace*{-\bigskipamount}
\end{algorithm}
\end{minipage}
\vspace{-\bigskipamount}
\end{figure}


\vspace{-\medskipamount}
\subsection{Detailed Algorithm Description}
\label{sub:erato:mw:desc}

To impose an ordering on the values written by the writers we 
associate each value with a tag $\tg{}$ defined as the pair $(ts,id)$, where 
$ts$ is a timestamp and $id$ the identifier of a writer.
Tags are ordered lexicographically (cf.~\cite{LS97}). 

\ppp{\em Reader Protocol.} 
Readers use state variables similarly to algorithm \algsw{}.
Reader $\rdr$ broadcasts a \textsf{readRequest}
message to all  servers, then awaits either 
(a) \textsf{readRelay} messages from some quorum, 
or 
(b) \textsf{readAck} messages from some quorum
%
(L\ref{line:erato:mw:bdcast:start}-\ref{line:erato:mw:bdcast:end}).
The key departure here
is in how the reader
handles case (a) when \vq{2} is detected, which indicates that the write
associated with the \emph{maximum} tag is not complete. 
	Here the reader considers past history 
 and discovers the tag associated with the last complete write.
	This is accomplished in an iterative manner, by removing  the 
	servers that respond with the maximum tag in the responding quorum $Q$ and repeating the analysis
	%
	(L\ref{line:erato:mw:reader:unravel:start}-\ref{line:erato:mw:reader:unravel:end}).
	During the iterative process, if $\rdr$ detects 
	\vq{1} it
	returns the value associated with the \emph{maximum} tag discovered 
	during the current iteration.
	If no iteration yields \vq{1}, 
	then eventually $\rdr$ observes \vq{3}. 
	In the last case, \vq{3} is detected when 
	a single server remains in some intersection of $Q$.
	If so, the reader waits \textsf{readAck} messages to arrive
	from some quorum, and returns the value associated with the \emph{minimum} tag. 
%
If case (b) happens before case (a), then $r$ proceeds identically 
as in the case where \vq{3} is detected
(L\ref{line:erato:mw:3exch:start}-\ref{line:erato:mw:3exch:end}).

\ppp{\em Writer Protocol.} 
Similarly to the four-exchange implementation \cite{LS97},
a writer
broadcasts a  {\sf writeDiscover} message to all servers, and
awaits  ``fresh'' \textsf{discoverAck} messages from some quorum  $Q$
(L\ref{line:erato:mw:writer:discover:start}-\ref{line:erato:mw:writer:discover:end}).
Among 
these responses
the writer finds the \emph{maximum} timestamp, $maxTS$,
increments it, and associates it and its own id with the new value
by broadcasting the new timestamp, its id, and the new value
in a \textsf{writeRequest} message to all servers.
The write completes when \textsf{writeAck} messages
are received from some quorum $Q$
(L\ref{line:erato:mw:writer:ph2:start}-\ref{line:erato:mw:writer:ph2:end}).

\ppp{\em Server Protocol.} 
Servers react to messages from  readers exactly as in Algorithm \ref{alg:erato-sw}.
We now describe hwo the messages from writers are handled.

$(1)$ Upon receiving message $\tup{{\sf writeDiscover},write\_op, w}$, 
 server $\srvr$ replies with a \textsf{discoverAck} message 
 containing   its local tag and value.
(L\ref{line:erato:mw:srv:discover:start}-\ref{line:erato:mw:srv:discover:end}).

$(2)$ Upon receiving message $\tup{{\sf writeRequest},ts^{\prime},id^{\prime},v^{\prime},write\_op, w}$, 
server $\srvr$ compares lexicographically its local tag
with the received one.
If $(ts,id) < (ts^{\prime},id^{\prime})$, then $\srvr$ 
updates its local information 
and replies using {\sf writeAck} message
%
(L\ref{line:erato:mw:srv:write-p2:start}-\ref{line:erato:mw:srv:write-p2:end}).

\vspace{\medskipamount}
\subsection{Correctness}
\label{sub:erato:mw:correct}

To prove correctness of algorithm \algmw{} 
we reason about its \emph{liveness} (termination) and \emph{atomicity} (safety)
as in Section \ref{sub:erato:sw:correct},
except using \emph{tags} instead of timestamps.

\paragraph{\bf\em Liveness.} 
Termination is satisfied with respect to our failure model:
at least one quorum $Q$ is non-faulty 
and each operation waits for messages from a quorum $Q$ of servers.
Let us consider this in more detail.
\smallskip

\emph{Write Operation.} 
Writer $w$ finds the maximum tag by broadcasting a \textsf{discover} message to all servers
and waits to collect \textsf{discoverAck} replies from a full quorum of servers 
(L\ref{line:erato:mw:writer:discover:start}-\ref{line:erato:mw:writer:discover:end}).
Since a full quorum of servers is non-faulty, then at least a full quorum of live servers 
will collect the \textsf{discover} messages and reply to writer $w$.
Once the maximum timestamp is determined, 
then writer $w$ updates its local tag
and broadcasts a \textsf{writeRequest} message to all servers.
Writer $w$ then waits to collect \textsf{writeAck} replies from a full quorum of servers before it terminates. 
Again, at a full quorum of servers will collect the \textsf{writeRequest} messages 
and will reply to writer $w$
(L\ref{line:erato:mw:writer:ph2:start}-\ref{line:erato:mw:writer:ph2:end}).
\smallskip 

\emph{Read Operation.} 
A read operation 
of the algorithm \algmw{} terminates when the reader $\rdr$ either
(i) collects \textsf{readAck} messages from full quorum of servers
or
(ii) collects \textsf{readRelay} messages from a full quorum and throughout
the iterative procedure it notices \vq{1} or \vq{3}
(L\ref{line:erato:mw:reader:unravel:start}-\ref{line:erato:mw:reader:unravel:end}).
Case (i) is identical as in Algorithm \algsw{} and
\emph{liveness} is ensured as reasoned in section \ref{sub:erato:sw:correct}.
For case (ii), in the worst case, during the iterative analysis 
\vq{3} will be noticed once one server remains in the replying quorum. 
This is identical to case (i) and the case follows. 

Based on the above, any read or write operation  collect a {sufficient} number of messages to terminate, guaranteeing \emph{liveness}.

\paragraph{\bf\em Atomicity.} 
As given in Section \ref{sub:erato:sw:correct}, atomicity can be reasoned about in terms of \emph{timestamps}. 
In the {\small MWMR} setting we use tags instead of timestamps, 
and here we show how algorithm \algmw{}
satisfies \emph{atomicity}  using \emph{tags}.
It is easy to see that the $\tg{}$ variable in each server $\srvr$ is monotonically 
increasing. This leads to the following lemma.


\begin{lemma} 
\label{lem:erato:mw:monotonicity}
In any execution $\EX$ of \algmw{}, the variable $\tg{}$ maintained by any server 
$\srvr$ in the system is non-negative and monotonically increasing.
\end{lemma}

\begin{Proof}
When server $\srvr$ receives a tag $\tg{}$ then $\srvr$ updates its 
local tag $\tg{s}$ if and only if $\tg{} > \tg{s}$ 
(L\ref{line:erato:mw:srv:relay-tg:start}-\ref{line:erato:mw:srv:relay-tg:end}, \ref{line:erato:mw:srv:relay-tag:start}-\ref{line:erato:mw:srv:relay-tag:end}). 

\end{Proof}


\begin{lemma}
\label{lem:erato:mw:write-after-read}
In any execution $\EX$ of \algmw{}, if a write $\wrt$ writes tag $\tg{}^{\prime}$ 
and succeeds a read operation $\rd$ that returns a tag $\tg{}$, i.e., 
 $\rd \bef \wrt$, 
then $\tg{}^{\prime} > \tg{}$.   
\end{lemma}

\begin{Proof}
Let $RR$ be the set of servers that belong to quorum $Q_a$ and sent \textsf{readRelay} messages to $\rho$.
Let $dAck$ be the set of servers from a quorum $Q_b$ that sent \textsf{discoverAck} messages to $\wrt$.
Let $wAck$ be the set of servers from a quorum $Q_c$ that sent \textsf{writeAck} messages to $\wrt$
and let $RA$ be the set of servers from a quorum $Q_d$ that sent \textsf{readAck} messages to $\rd$. 
It is not necessary that $a \neq b \neq c \neq d$ holds.

Based on the read protocol, 
the read operation $\rd$ terminates when it either receives
(a) \textsf{readRelay} messages from a full quorum $Q$ 
or 
(b) \textsf{readAck} messages from a full quorum $Q$
(L\ref{line:erato:mw:bdcast:start}-\ref{line:erato:mw:bdcast:end}).

Case (a), 
based on the algorithm, during the iterative analysis 
$\rho$ terminates once it notices \vq{1} or \vq{3} in the messages received from $RR$.
If \vq{1} is noticed then the distribution of the timestamps indicates the 
existence of one and only tag in $Q_a$ and that is, the \emph{maximum} tag
in $Q_a$ at the current iteration.
Read $\rho$ returns the value associated with the current \emph{maximum} tag, $\tg{}$
and terminates.
The following
writer $\wrt$, initially it broadcasts a \textsf{writeDiscover} message to all servers, and
it then awaits for ``fresh'' \textsf{discoverAck} messages from a full quorum $Q_b$, that is, set $dAck$
(L\ref{line:erato:mw:writer:discover:start}-\ref{line:erato:mw:writer:discover:end}).
Observe that each of $RR$ and $dAck$ sets are from a full quorum of servers, $Q_a$ and $Q_b$ respectivelly, 
and so $RR\cap dAck\neq\emptyset$. 
By Lemma \ref{lem:erato:mw:monotonicity}, any server $\srvr_k \in RR\cap dAck$ has a tag 
$\tg{\srvr_k}$ s.t. $\tg{\srvr_k} \geq \tg{}$. 
Since $\wrt$ generates a new local tag-value $(\tg{}^{\prime},v)$ pair
in which it assigns the timestamp to be one higher than the one discovered in the \emph{maximum} tag
from set $dAck$,
it follows that $\tg{}^{\prime} > \tg{}$.   
Write operation $\wrt$ broadcasts the value to be written associated with $\tg{}^{\prime} $ 
in a \textsf{writeRequest} message to all servers 
and it awaits for \textsf{writeAck} messages from a full quorum $Q_c$ before completion, set $wAck$
(L\ref{line:erato:mw:writer:ph2:start}-\ref{line:erato:mw:writer:ph2:end}).
Observe that each of $dAck$ and $wAck$ sets are from a full quorum of servers, $Q_b$ and $Q_c$ respectivelly, 
and so $dAck\cap wAck\neq\emptyset$. 
By Lemma \ref{lem:erato:mw:monotonicity}, any server $\srvr_k \in dAck\cap wAck$ has a tag 
$\tg{\srvr_k}$ s.t. $\tg{\srvr_k} \geq \tg{}^{\prime} > \tg{} $ and the result
for this case follows. 

Now we examine if \vq{3} is noticed.
When that holds, based on the algorithm, 
the reader awaits \textsf{readAck} messages from a full quorum $Q$ of servers, set $RA$.
By lines 
\ref{line:erato:mw:3exch:start} - \ref{line:erato:mw:3exch:end}, 
it follows that $\rd$ decides on the minimum tag, $\tg{}=minTG$, 
among all the tags in the \textsf{readAck} messages of the set $RA$
and terminates.
Again, $\wrt$, initially it broadcasts a \textsf{writeDiscover} message to all servers, and
it then awaits for ``fresh'' \textsf{discoverAck} messages from a full quorum $Q_b$, that is, set $dAck$.
Each of $RA$ and $dAck$ sets are from a full quorum of servers, $Q_d$ and $Q_b$ respectivelly, 
and so $RA\cap dAck\neq\emptyset$. 
By Lemma \ref{lem:erato:mw:monotonicity}, any server $\srvr_k \in RA\cap dAck$ has a tag 
$\tg{\srvr_k}$ s.t. $\tg{\srvr_k} \geq \tg{}$. 
Since $\wrt$ generates a new local tag-value $(\tg{}^{\prime},v)$ pair
in which it assigns the timestamp to be one higher than the one discovered in the \emph{maximum} tag
from set $dAck$,
it follows that $\tg{}^{\prime} > \tg{}$.  
Furthermore, $\wrt$ broadcasts the value to be written associated with $\tg{}^{\prime} $ 
in a \textsf{writeRequest} message to all servers 
and it awaits for \textsf{writeAck} messages from a full quorum $Q_c$ before completion, set $wAck$
(L\ref{line:erato:mw:writer:ph2:start}-\ref{line:erato:mw:writer:ph2:end}).
Observe that each of $dAck$ and $wAck$ sets are from a full quorum of servers, $Q_b$ and $Q_c$ respectivelly, 
and so $dAck\cap wAck\neq\emptyset$. 
By Lemma \ref{lem:erato:mw:monotonicity}, any server $\srvr_k \in dAck\cap wAck$ has a tag 
$\tg{\srvr_k}$ s.t. $\tg{\srvr_k} \geq \tg{}^{\prime} > \tg{} $ and the result
for this case follows. 

Lastly, case (b) where read $\rho$ terminates because it received \textsf{readAck} messages 
from a full quorum of servers $Q$, it is the same as in case (a) when reader observers \vq{3}
and the lemma follows. 
\end{Proof}


\begin{lemma}
\label{lem:erato:mw:write-after-write}
In any execution $\EX$ of \algmw{}, 
if a write $\omega_1$ writes 
tag $\tg{1}$ and precedes a write
$\omega_2$ that writes 
tag $\tg{2}$, i.e., $\omega_1 \to \omega_2$,
then $\tg{2} > \tg{1}$. 
\end{lemma}

\begin{Proof}
Let $wAck_1$ be the set of servers from a full quorum $Q_a$ that send a \textsf{writeAck} message 
within write operation $\omega_1$. Let $dAck_2$ be the set of servers from a full quorum $Q_b$ (not necessarily different from $Q_a$)
that send a \textsf{discoverAck} message within  write operation $\omega_2$.

Lemma assumes that $\omega_1$ is complete. 
By Lemma \ref{lem:erato:mw:monotonicity}, we know that if a server $\srvr$ receives a tag $\tg{}$ from a process $\pr$,
then $\srvr$  includes tag $\tg{}^{\prime}$ s.t. $\tg{}^{\prime} \geq \tg{}$ in any subsequent message.
Thus, servers in $wAck_1$ send a \textsf{writeAck} message within $\omega_1$ 
with tag at least tag $\tg{1}$. 
%

Once $\omega_2$ is invoked, 
it
collects \textsf{discoverAck} messages from a full quorum of servers in the set, $dAck_2$
(L\ref{line:erato:mw:writer:discover:start} -\ref{line:erato:mw:writer:discover:end}).
Since $Q_a \subseteq wAck_1$ and $Q_b \subseteq dAck_2$
%
then $wAck_1\cap dAck\neq\emptyset$. 
By Lemma \ref{lem:erato:mw:monotonicity}, any server $\srvr_k \in wAck_1\cap dAck_2$ has a tag 
$\tg{\srvr_k}$ s.t. $\tg{\srvr_k} \geq \tg{1}$. 
Thus, the invoker of $\omega_2$ discovers the maximum tag, $maxTG$, from the tags found in $dAck_2$
s.t. $maxTG \geq \tg{\srvr_k} \geq \tg{1}$
(L\ref{line:erato:mw:writer:find-max}).
It then increases the timestamp from in the maximum tag discovered by one, 
sets it's local tag to that and associates it with its id $i$ and the value $val$ to be written,
$local\_tag = (maxTS+1, i, val)$
(L\ref{line:erato:mw:writer:create-tag}).
We know that, $local\_tag > maxTG \geq \tg{1}$, hence $local\_tag > \tg{1}$. 

Lastly, $\omega_2$ attaches its local tag 
$local\_tag$ in a \textsf{writeRequest} message
which it broadcasts to all the servers, 
and terminates upon receiving \textsf{writeAck} messages 
from a full quorum of servers.
By Lemma \ref{lem:erato:mw:monotonicity}, $\omega_2$ receives \textsf{writeAck} 
messages with a tag $\tg{2}$ s.t. $\tg{2} \geq local\_tag > \tg{1}$ hence $\tg{2} > \tg{1}$. 
This completes the Proof of the lemma.~\vspace{.3em}
\end{Proof}


We now show that any read operation that follows a write operation,
and it receives \textsf{readAck} messages the servers where each included tag 
is at least as the one returned by the complete write operation.

\begin{lemma}
\label{lem:erato:mw:read-received-timestamps}
In any execution $\EX$ of \algmw{}, if a read operation $\rd$ succeeds a write 
operation $\omega$ that writes $\tg{}$ and $v$, i.e., $\omega \bef \rd$, and 
receives \textsf{readAck} messages from a quorum $Q$ of servers, set $RA$, then 
each $\srvr\in RA$ sends a \textsf{readAck} message to $\rd$ with a 
tag $\tg{s}$ s.t. $\tg{s} \geq \tg{}$. 
\end{lemma}

\begin{Proof}
Let $wAck$ be the set of servers from a quorum $Q_a$ that send a \textsf{writeAck} message to $\omega$, 
let $RelaySet$ be the set of servers from a quorum $Q_b$ that sent \textsf{readRelay} messages to server $\srvr$,
and let $RA$ be the set of servers from a quorum $Q_c$ that send a \textsf{readAck} message to $\rd$. 
Notice that it is not necessary that $a \neq b \neq c$ holds.

Write operation $\omega$ is completed. 
By Lemma \ref{lem:erato:mw:monotonicity}, if a server $\srvr$ receives a tag $\tg{}$ 
from a process $\pr$ at some time $t$, then $\srvr$ attaches a tag $tg{}^{\prime}$ 
s.t. $\tg{}^{\prime} \geq ts$ in any message sent at any time $t^{\prime} \geq t$. 
Thus,
every server in $wAck$, sent a \textsf{writeAck} message to $\omega$ 
with a tag greater or equal to $\tg{}$. 
Hence, every server $\srvr \in wAck$ has a tag $\tg{\srvr} \geq \tg{}$. 
Let us now examine a tag $\tg{s}$ that server $\srvr$ sends to read operation $\rd$.  

Before server $\srvr$ sends a \textsf{readAck} message to $\rd$, 
it must receive \textsf{readRelay} messages from a full quorum $Q_b$ of servers, $RelaySet$
(L\ref{line:erato:mw:srv:read-ack:start}-\ref{line:erato:mw:srv:read-ack:end}).
Since both $wAck$ and $RelaySet$ contain messages from a full quorum of servers, 
and by definiton, any two quorums have a non-empty intersection, then $wAck\cap RelaySet\neq\emptyset$. 
By Lemma \ref{lem:erato:mw:monotonicity}, any server $\srvr_x \in aAck \cap RelaySet$ has a tag $\tg{s_x}$ s.t. $\tg{s_x} \geq \tg{}$.
Since server $\srvr_x \in RelaySet$ and from the algorithm, 
server's $\srvr$ tag is always updated to the highest tag it noticed 
(L\ref{line:erato:mw:srv:relay-tag:start}-\ref{line:erato:mw:srv:relay-tag:end}),
then when server $\srvr$ receives the message from $\srvr_x$, 
it will update its tag $\tg{s}$ s.t. $\tg{s} \geq \tg{s_x}$.
Server $\srvr$ creates a \textsf{readAck} message where it encloses 
its local tag and its local value, $(\tg{s}, v_s)$
(L\ref{line:erato:mw:srv:send:readack}).
Each $\srvr \in RA$ sends a \textsf{readAck} to $\rd$ with a tag 
$\tg{s}$ s.t. $\tg{s} \geq \tg{s_x} \geq \tg{}$. 
Thus, $\tg{s} \geq \tg{}$, and the lemma follows.
\end{Proof}


Now, we show that if a read operation succeeds a write operation, 
then it returns a value at least as recent as the one written.\vspace{.5em}

\begin{lemma}
\label{lem:erato:mw:read-after-write}
In any execution $\EX$ of \algmw{}, if a read $\rd$ succeeds a write operation $\omega$ 
that writes tag $\tg{}$, i.e. $\omega \to \rd$, and returns a tag $\tg{}^{\prime}$, then $\tg{}^{\prime} \geq \tg{}$.
\end{lemma}

\begin{Proof}
A read operation $\rd$ terminates when it either receives
(a) \textsf{readRelay} messages from a full quorum $Q$ 
or 
(b) \textsf{readAck} messages from a full quorum $Q$
(L\ref{line:erato:mw:bdcast:start}-\ref{line:erato:mw:bdcast:end}).

We first examine case (b). 
Let's suppose that $\rd$ receives \textsf{readAck} messages from a full quorum $Q$ of servers, $RA$. 
By lines 
\ref{line:erato:mw:3exch:start} - \ref{line:erato:mw:3exch:end}, 
it follows that $\rd$ decides on the minimum tag, $\tg{}^{\prime}=minTG$, 
among all the tag in the \textsf{readAck} messages of the set $RA$.
From Lemma \ref{lem:erato:mw:read-received-timestamps}, $minTG \geq \tg{}$ holds, 
where $\tg{}$ is the tag written by the last complete write operation $\omega$. 
Then $\tg{}^{\prime} = minTG \geq \tg{}$ also holds. 
Thus, $\tg{}^{\prime} \geq \tg{}$.

Now we examine case (a). 
Case (a) is an iterative procedure that terminates when the reader notices either
(i) \vq{1} 
or 
(iii) \vq{3}.
When \vq{2} is observed then it is the case where
the write associated with the \emph{maximum} tag is not yet 
complete, thus we proceed to the next iteration to discover
the latest potentially complete write. This, by removing
all the servers with the \emph{maximum} tag from $Q$ and repeating the analysis. 
If no iteration was interrupted because of \vq{1} then
eventually \vq{3} will be noticed, when a single server
remains in some intersection of $Q$ (L\ref{line:erato:mw:reader:unravel:start}-\ref{line:erato:mw:reader:unravel:end}).

Let $wAck$ be the set of servers from a quorum $Q_a$ that send a \textsf{writeAck} message to $\omega$.
Since the write operation $\omega$, that wrote value $v$ associated with tag $\tg{}$ is complete,
and by monotonicity of tags in servers (Lemma \ref{lem:erato:mw:monotonicity}),  
then at least a quorum $Q_a$ of servers has a tag $\tg{a}$ s.t. $\tg{a} \geq \tg{}$.

Let's suppose that $\rd$ receives \textsf{readRelay} messages from a full quorum $Q_b$ of servers, $RR$. 
Since both $wAck$ and $RR$ contain messages from a full quorum of servers, quorums $Q_a$ and $Q_b$,
and by definition any two quorums have a non-empty intersection, then $wAck \cap RR \neq \emptyset$. 
Since every server in $wAck$ has a tag $\tg{a} \geq \tg{}$ then 
any server $s_x \in wAck \cap RR$ has a tag $\tg{s_x}$ s.t. $\tg{s_x} \geq \tg{a} \geq \tg{}$.

Assume by contradiction that at the $i^{th}$ iteration $\rd$ noticed \vq{1} in $RR$
and returned a tag $\tg{}^{\prime}$ s.t. $\tg{}^{\prime} < \tg{}$.
Since every server $s_x \in wAck \cap RR$ has a tag $\tg{s_x}$ s.t. $\tg{s_x} \geq \tg{}$
and since \vq{1} returned a tag $\tg{}^{\prime}$ s.t. $\tg{}^{\prime} < \tg{}$,
then it must be the case that none of the servers in $wAck \cap RR$ were
participating in \vq{1}. 
Therefore, it must be the case that all servers in $wAck \cap RR$ 
were removed during the analysis at a previous iteration $k$, s.t. $k<i$.
However, we know that the iterative procedure, in the worst case, it will 
notice \vq{3} once a single server remains in an intersection 
of the quorum we examine.
This contradicts the fact that all servers in $wAck \cap RR$ were removed 
from $Q_a$ during the analysis. Thus, if \vq{1} is noticed, then 
the distribution of the tags yelled the existence of one and only tag,
the current \emph{maximum} tag. At least one server $s_x$ from $\in wAck \cap RR$
will participate in \vq{1}, hence
$\rd$ will return a tag $\tg{}^{\prime}$ s.t. $\tg{}^{\prime} = \tg{s_x} \geq \tg{}$.

Lastly, when \vq{3} is noticed during the iterative procedure then $\rd$
waits for \textsf{readAck} messages from a full quorum $Q$ before termination, 	
(L\ref{line:erato:mw:reader:qview3:start}-\ref{line:erato:mw:reader:qview3:end}),
proceeds identically as in case (b) and the lemma follows.
\end{Proof}


\begin{lemma}
\label{lem:erato:mw:semifast-read-after-read}
In any execution $\EX$ of \algmw{}, if $\rd_1$ and $\rd_2$ 
are two \emph{semi-fast} read operations, take 3 exchanges to complete, 
such that $\rd_1$ precedes $\rd_2$, i.e., $\rd_1 \to \rd_2$, 
and $\rd_1$ returns the value for tag $\tg{1}$, then $\rd_2$ returns 
the value for tag $\tg{2} \geq \tg{1}$.
\end{lemma}

\begin{Proof} 
Let the two operations $\rd_1$ and $\rd_2$ be invoked by processes with identifiers
$\rdr_1$ and $\rdr_2$ respectively (not necessarily different). Also, let $RA_1$ and 
$RA_2$ be the sets of servers from quorums $Q_a$ and $Q_b$ (not necessarily different)
that sent a \textsf{readAck} message to $\rdr_1$ and 
$\rdr_2$ during $\rd_1$ and $\rd_2$.

Assume by contradiction
that read operations $\rd_1$ and $\rd_2$ exist such that $\rd_2$ succeeds $\rd_1$, 
i.e., $\rd_1 \to \rd_2$, and the operation $\rd_2$ returns a tag $\tg{2}$ 
that is smaller than the $\tg{1}$ returned by $\rd_1$, i.e., $\tg{2} < \tg{1}$. 
Based on the algorithm,  
$\rd_2$ returns a tag $\tg{2}$ that is smaller than 
the minimum tag received by $\rd_1$, i.e., $\tg{1}$, 
if $\rd_2$ obtains $\tg{2}$ and $v$ in the \textsf{readAck} 
message of some server $\srvr_x\in RA_2$, 
and $\tg{2}$ is the minimum tag received by $\rd_2$.

Let us examine if $\srvr_x$  
sends a \textsf{readAck} message to $\rd_1$ with tag $\tg{x}$,
i.e., $\srvr_x\in RA_1$. 
By Lemma \ref{lem:erato:mw:monotonicity}, and since $\rd_1\bef\rd_2$, then 
it must be the case that $\tg{x} \leq \tg{2}$.
According to our assumption $\tg{1} > \tg{2}$, and since $\tg{1}$ is the 
smallest tag sent to $\rd_1$ by any server in $RA_1$, then it follows that $\rdr_1$ does 
not receive the \textsf{readAck} message from $\srvr_x$, and hence $\srvr_x\notin RA_1$. 

Now let us examine the actions of the server $\srvr_x$. 
From the algorithm, server $\srvr_x$ collects \textsf{readRelay} messages from a full quorum $Q_c$
of servers before sending a \textsf{readAck} message to $\rd_2$
(L\ref{line:erato:mw:srv:relay-ack:end}-\ref{line:erato:mw:srv:relay-ack:end}).  
Let $RRSet_{\srvr_x}$ be the set of servers that belong to quorum $Q_c$ 
and sent \textsf{readRelay} message to $\srvr_x$.
Since, both $RRSet_{\srvr_x}$ and $RA_1$ contain messages from full quorums, $Q_c$ and $Q_a$,
and since any two quorums have a non-empty intersection, 
then it follows that $RRSet_{\srvr_x}\cap RA_1\neq\emptyset$.

Thus there exists a server $\srvr_i \in RRSet_{\srvr_x}\cap RA_1$, 
that sent 
(i) a \textsf{readAck} to $\rd_1$, 
and 
(ii) a \textsf{readRelay} to $\srvr_x$ during $\rd_2$. 
Note that $\srvr_i$ sends a \textsf{readRelay} for $\rd_2$ only after
it receives a read request from $\rd_2$. 
Since $\rd_1\bef \rd_2$, then it follows that $\srvr_i$ sent the 
\textsf{readAck} to $\rd_1$ before sending the \textsf{readRelay} to $\srvr_x$. 
By Lemma \ref{lem:erato:mw:monotonicity}, if $\srvr_i$ attaches a tag $\tg{s_i}$ 
in the \textsf{readAck} to $\rd_1$, then $\srvr_i$ attaches a tag $\tg{s_i}^{\prime}$ 
in the \textsf{readRelay} message to $\srvr_x$, such that  $\tg{s_i}^{\prime} \geq \tg{s_i}$. 
Since $\tg{1}$ is the minimum tag received by $\rd_1$, 
then $\tg{s_i} \geq \tg{1}$, and hence $\tg{s_i}^{\prime} \geq \tg{1}$ as well. 
By Lemma \ref{lem:erato:mw:monotonicity}, and since $\srvr_x$ receives the \textsf{readRelay} 
message from $\srvr_i$ before sending a \textsf{readAck} to $\rd_2$, 
it follows that $\srvr_x$ sends a tag $\tg{2}$ to $\rho_2$ s.t. $\tg{2} \geq \tg{s_i}^{\prime} \geq \tg{1}$.
Thus, $\tg{2} \geq \tg{1}$ and this contradicts our initial assumption. 
\end{Proof}


\begin{lemma}
\label{lem:erato:mw:fast-read-after-read}
In any execution $\EX$ of \algmw{}, if $\rd_1$ and $\rd_2$ 
are two \emph{fast} read operations, take 2 exchanges to complete, 
such that $\rd_1$ precedes $\rd_2$, i.e., $\rd_1 \to \rd_2$, 
and $\rd_1$ returns the value for tag $\tg{1}$, then $\rd_2$ returns 
the value for tag $\tg{2} \geq \tg{1}$.
\end{lemma}

\begin{Proof} 
Let the two operations $\rd_1$ and $\rd_2$ be invoked by processes with identifiers
$\rdr_1$ and $\rdr_2$ respectively (not necessarily different). 
Let $RR_1$ and $RR_2$ be the quorums (not necessarily different)
that sent a \textsf{readRelay} message to $\rdr_1$ and 
$\rdr_2$ during $\rd_1$ and $\rd_2$ respectively.

The algorithm terminates in \emph{two} communication exchanges
when a read operation $\rd$ receives \textsf{readRelay} messages 
from a full quorum $Q$ and based on the distribution of the tags
during the $i^{th}$ iteration of the analysis, $i\geq 1$, it notices \vq{1}.

Observe that if there exists a server $s_k \in RR_1$, 
that replies with a tag $\tg{s_k}$ s.t. $\tg{s_k} < \tg{1}$
then $\rd_1$ wouldn't be able to notice \vq{1} and return $\tg{1}$.
Thus, since \vq{1} is noticed during the $i^{th}$ iteration then
it is known that all the servers in $RR_1$ replied to $\rd_1$ with a tag 
$\tg{s}$ s.t. $\tg{s} \geq \tg{1}$.
This is clear since every server $s_x$ that was removed during the iterative analysis at iteration $j$ 
s.t. $j > i$, server $s_x$ holds a tag $\tg{s_x} \geq \tg{s}$.

Since by definition, any two quorums have a non-empty intersection
it follows that $RR_1 \cap RR_2 \neq \emptyset$.
From that and by Lemma \ref{lem:erato:mw:monotonicity}, then every server
$s_x \in RR_1 \cap RR_2$ has a tag $\tg{}^{\prime}$ such that
$\tg{}^{\prime} \geq \tg{1}$.
When $\rd_2$ notices \vq{1} in $RR_2$ at the $m^{th}$ iteration of the analysis, $m\geq 1$, 
we know that \vq{1} consists tags that come from the set $RR_1 \cap RR_2$.
Notice that if $RR_1 \cap RR_2 = \emptyset$ holds at iteration $m$, 
then it means that the algorithm would have stopped at an earlier iteration 
when either $RR_1 \cap RR_2 \neq \emptyset$
or $|RR_1 \cap RR_2| = 1$ holds.

Since the distribution of the tags during $m^{th}$ iteration indicates the existence of one 
and only tag and since servers from $RR_1 \cap RR_2$ participate 
then $\rd$ returns a value associated with $\tg{2}$ s.t. $\tg{2} \geq \tg{}^{\prime} \geq \tg{1}$
and the lemma follows.
\end{Proof}


\begin{lemma}
\label{lem:erato:mw:read-after-read}
In any execution $\EX$ of \algmw{}, if $\rd_1$ and $\rd_2$ 
are two read operations
s.t. $\rd_1$ precedes $\rd_2$, i.e., $\rd_1 \to \rd_2$, 
and $\rd_1$ returns 
tag $\tg{1}$, then $\rd_2$ returns 
a tag $\tg{2}$, s.t. $\tg{2} \geq \tg{1}$.
\end{lemma}

\begin{Proof} 
We are interested to examine the cases where one of the read operation is 
\emph{fast} and the other is \emph{semifast}.
In particular, cases
(i) $\rd_1 \to \rd_2$ and $\rd_1$ is \emph{semifast} and $\rd_2$ is \emph{fast}
and
(ii) $\rd_1 \to \rd_2$ and $\rd_1$ is \emph{fast} and $\rd_2$ is \emph{semifast}.

Let the two operations $\rd_1$ and $\rd_2$ be invoked by processes with identifiers
$\rdr_1$ and $\rdr_2$ respectively (not necessarily different). 
Also, let $RR_1$,$RA_1$ and $RR_2$,$RA_2$ be the sets of servers 
from full quorums (not necessarily different)
that sent a \textsf{readRelay} and \textsf{readAck} message to $\rd_1$ and $\rd_2$ respectively.

We start with case (i). 
Since read operation $\rd_1$ is \emph{semifast}, then based on the algorithm,
the tag $\tg{1}$ that is returned it is also the \emph{minimum} tag noticed
in $RA_1$.
Before a server $s$ sents \textsf{readAck} messages to $\rd_1$ (that form $RA_1$),
it must receive \textsf{readRelay} messages from a full quorum of servers.
Thus, by Lemma \ref{lem:erato:mw:monotonicity} monotonicity of the tags at the servers
we know that the \emph{minimum} tag that a full quorum has by the end
of $\rd_1$ is $\tg{1}$.
Read operation $\rd_2$ receives \textsf{readRelay} messages from a full quorum of servers, $RR_2$.
By definition of quorums, since both $RA_1$ and $RR_2$ are from a full quorum of
servers then it follows that $RA_1 \cap RR_2 \neq \emptyset$.
Thus every server $s_x \in RA_1 \cap RR_2$ holds a tag $\tg{}^{\prime}$ s.t. $\tg{}^{\prime} \geq \tg{1}$.

For $\rd_2$ to notice \vq{1} in $RR_2$ at the $m^{th}$ iteration of the analysis, $m\geq 1$, 
it means that in \vq{1} participate servers that belong to $RR_1 \cap RR_2$.
Notice that if $RR_1 \cap RR_2 = \emptyset$ holds at iteration $m$, 
then it means that the algorithm would have stopped at an earlier iteration 
when either $RR_1 \cap RR_2 \neq \emptyset$
or $|RR_1 \cap RR_2| = 1$ holds.
Since the distribution of the tags during $m^{th}$ iteration indicates the existence of one 
and only tag and since servers from $RR_1 \cap RR_2$ participate 
then $\rd$ returns a value associated with $\tg{2}$ s.t. $\tg{2} \geq \tg{}^{\prime} \geq \tg{1}$
and the case follows.

We now examine case (ii). 
Since $\rd_1$ is fast, it follows that it has noticed \vq{1} in $RR_1$. 
If \vq{1} was noticed at the $m^{th}$ iteration of the analysis, $m\geq 1$, 
and $\rd_1$ returned a value associated with \emph{maximum} tag during $m^{th}$ iteration, $\tg{1}$, 
then by the completion of $\rd_1$
a full quorum has a tag $\tg{}^{\prime}$ s.t. $\tg{}^{\prime} \geq \tg{1}$.
Now, since read operation $\rd_2$ is \emph{semifast}, then based on the algorithm,
the tag $\tg{2}$ that is returned it is the \emph{minimum} tag noticed
in $RA_2$.
Before a server $s$ sents \textsf{readAck} messages to $\rd_2$ (that form $RA_2$),
it must receive \textsf{readRelay} messages from a full quorum of servers, $RelaySet$.
By Lemma \ref{lem:erato:mw:monotonicity} monotonicity of the tags at the servers
and $RR_1 \cap RelaySet \neq \emptyset$, then every server in $RA_2$
has a tag $\tg{2}$ s.t. $\tg{2} \geq \tg{}^{\prime} \geq \tg{1}$ and the case follows.

The rest of the cases are proved in Lemmas \ref{lem:erato:mw:semifast-read-after-read} and \ref{lem:erato:mw:fast-read-after-read}
and the lemma follows.
\end{Proof}

\begin{theorem}
Algorithm {\small \sc \algmw{}} implements an atomic {\small MWMR} object.
\end{theorem}

\begin{Proof} 
We use the above lemmas and the operations order definition 
to reason about each of the three {atomicity} conditions A1, A2 and A3. 

\ppp{A1} For any $\op_1,\op_2\in\Pi$ such that $\op_1\bef\op_2$, it cannot be that $\op_2\prec\op_1$.

\noindent	If both $\op_1$ and $\op_2$ are writes and $\op_1\bef\op_2$ holds,
				then from Lemma~\ref{lem:erato:mw:write-after-write}
				it follows that 
				$\tg{\op_2} > \tg{\op_1}$.
				From the definition of order $\prec$ we have $\op_1\prec\op_2$.
				When $\op_1$ is a write, $\op_2$ a read and $\op_1\bef\op_2$ holds,
				then from Lemma~
				\ref{lem:erato:mw:read-after-write}
				it follows that 
				$\tg{\op_2} \geq \tg{\op_1}$.
				By definition $\op_1\prec\op_2$ holds.
				If $\op_1$ is a read, $\op_2$ is a write and $\op_1\bef\op_2$ holds,
				then from Lemma~\ref{lem:erato:mw:write-after-read}
				it follows that $\op_2$  returns a tag $\tg{\op_2}$ s.t. $\tg{\op_2} > \tg{\op_1}$.
				By the order definition $\op_1\prec\op_2$ is satisfied.		
				If both $\op_1$ and $\op_2$ are reads and $\op_1\bef\op_2$ holds, then	
				from Lemma~
				\ref{lem:erato:mw:read-after-read}
				it follows that 
				$\tg{\op_2} \geq \tg{\op_1}$.
				If $\tg{\op_2} > \tg{\op_1}$, then by the ordering definition $\op_1\prec\op_2$ holds.
				When $\tg{\op_2} = \tg{\op_1}$ then the ordering is not defined, thus it cannot be that
				$\op_2\prec\op_1$.

\ppp{A2} For any write $\wrt\in\Pi$ and any operation $\op\in\Pi$, then either $\wrt\prec\op$ or $\op\prec\wrt$.
			
\noindent 	If $\tg{\wrt} > \tg{\op}$, then $\op\prec\wrt$ follows directly.
				Similarly, if $\tg{\wrt} < \tg{\op}$ holds, then it follows that $\wrt\prec\op$.
				When $ts_{\wrt} = ts_{\pi}$ holds, then because all writer tags are unique
				(each server increments timestamps monotonically, and the server ids
				disambiguate among servers)
				$\op$ can only be a read.
%
				%
				Since $\op$ is a read and the distribution of the tag written by $\wrt$ 
				satisfies either \vq{1} or \vq{3}, it follows that $\wrt\prec\op$.

\ppp{A3} Every read operation returns the value of the last write preceding it according to $\prec$ (or the initial value if there is no such write).
	
\noindent	Let $\wrt$ be the last write preceding read $\rho$.
				From our definition it follows that $\tg{\rho} \geq \tg{\wrt}$.
		  		If $\tg{\rho} = \tg{\wrt}$, then $\rho$ returned
		  	  	a value written by $\wrt$ in some servers in a quorum $Q$. 
		    		Read $\rho$ either was \emph{fast} and during the iterative analysis it noticed
		    		a distribution of the tags in $Q$ that satisfied \vq{1}
		    		or $\rho$ was \emph{slow} and waited for \textsf{readAck} messages from a full quorum $Q$.
		   		 In the latter, the intersection properties of quorums ensure that $\wrt$ was the last complete write.    
		   		 %
		    		%
		    		If $\tg{\rho} > \tg{\wrt}$ holds, it must be the case that there is a write $\wrt'$ 
		    		that wrote $\tg{\rho}$ and by definition it must hold that $\wrt \prec \wrt^{\prime}\prec\rd$.
		    		Thus, $\wrt$ is not the preceding write and this cannot be the case.
		    		Lastly, if $\tg{\rho} = 0$, no preceding writes exist, and $\rho$ returns the initial value.
\end{Proof}

Proving liveness and atomicity of algorithm \algmw{} the result follows.

\subsection{Performance}
\label{sub:erato:mw:performance}

By inspection of the code, write operations take {2} \metrics{} and
read operations take either {2} or {3} \metrics{}.
The (worst case) message complexity of 
write operations is $2|\srvSet|$
and of read operations
is $|\srvSet|^{2} + 2|\srvSet|$,
as follows from the structure of the algorithm.
We now provide additional details.

\ppp{Operation Latency.}
\emph{Write operation latency:} 
According to algorithm \algmw{}, writer $w$ sends \textsf{discover} messages 
to all servers in exchange \x{1}
and waits for \textsf{discoverAck} messages from a full quorum of servers in \x{2}.
Once the \textsf{discoverAck} messages are received from \x{2}, then writer $w$ broadcasts 
a \textsf{writeRequest} message 
to all servers in \x{3}. 
It then waits for \textsf{writeAck} messages from a full quorum of servers from \x{4}.
No further communication is required and the write operation terminates. 
Thus a write operation  consists of \emph{4} communication exchanges.

\emph{Read operation latency:} 
A reader sends a {\sf{readRequest}} message to all the servers
in the first communication exchange \x{1}. 
Once the servers receive the \textsf{readRequest}
message they broadcast a \textsf{readRelay} message to 
all servers \emph{and} the reader in exchange \x{2}.
The reader can terminate at the end of the \x{2}
if it receives \textsf{readRelay} messages and based on the distribution of 
the tags through the iterative procedure it notices \vq{1} or \vq{3}.
If not, the operation goes into the third exchange \x{3}.
Thus read operations terminate after either \emph{2} or \emph{3} communication exchanges.

\ppp{Message Complexity.}
\emph{Write operation:} 
Write operations in algorithm \algmw{}
take \emph{4} communication exchanges. 
The first and the third exchanges, \x{1} and \x{3}, occur when a writer sends \textsf{discover} and \textsf{writeRequest} messages respectively to all servers in $\srvSet$. 
The second and fourth exchanges, \x{2} and \x{4}, occur when servers in $\srvSet$ send \textsf{discoverAck} and \textsf{writeAck} messages respectively back to the writer. 
Thus $4|\srvSet|$ messages are exchanged in any write operation. 

\emph{Read operation:} 
The structure of the read protocol of \algmw{} is similar as in \algsw{}, thus,
as reasoned in Section~\ref{sub:erato:sw:performance},
$|\srvSet|^{2} + 3|\srvSet|$ messages are exchanged during a read operation.

%% file: simulation.tex
We now compare the algorithms using the NS3 discrete event simulator \cite{NS3}. 
The following {\small SWMR} algorithms
\algsw{}, \ABD{} \cite{ABD96}, \osam{} \cite{HNS17}, and \sliq{} \cite{GNS08},
and the corresponding {\small MWMR} algorithms:
\algmw{}, \ABDmwmr{} \cite{LS97}, 
\omam{} \cite{HNS17}, and \cwfr{} \cite{GNRS11} were simulated.
For comparison, we implemented benchmark \swimp{}
that mimics the minimum message requirements:
%
\swimp{} does two exchanges for reads and writes, 
and neither performs any computation nor ensures consistency.\vspace{1mm} 
%

\begin{wrapfigure}{R}{0.6\textwidth}
\vspace{-\bigskipamount}
\includegraphics[width= 0.6\textwidth]{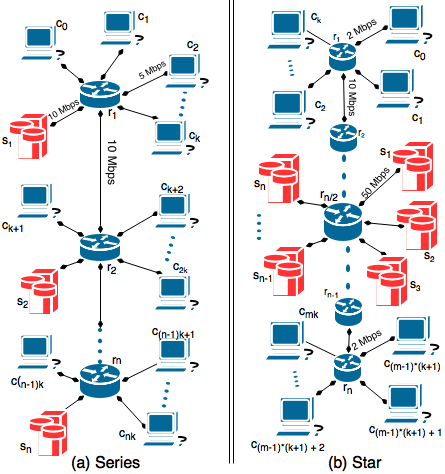}
\vspace{-\bigskipamount}
\caption{Simulated topologies.}
\label{fig:erato:topologies}
\vspace{-\bigskipamount}
\end{wrapfigure}

We developed two
topologies
that use the same array of routers,
but differ in the deployment of server and client nodes.
Clients are connected to routers over 5Mbps links
with 4ms delay and the routers 
over 10Mpbs links with 6ms delay. 
In \emph{Series} topology, Fig.\ref{fig:erato:topologies}(a),
a server is connected to each router over 10Mbps bandwidth with 2ms delay,
modeling a network where servers are separated and appear to be in different networks. 
In \emph{Star} topology, Fig.\ref{fig:erato:topologies}(b),
servers are connected to a single router over 50Mbps links with 2ms delay,
modeling a network where servers are in close proximity and well-connected,
e.g., a datacenter.
%
Clients are located uniformly
with respect to the routers.
%
%

\noindent{\bf Performance.}
We assess algorithms in terms of \emph{operation latency} that depends on 
communication delays and local computation time.
%
For operation latency 
we combine two clocks: 
the simulation clock to measure communication delays, and 
a real time clock for computation delays.
The sum of the two yields latency.

\noindent{\bf Experimentation Setup.}
To subject the system to high communication traffic, no failures are assumed
(ironically, crashes reduce network traffic). 
Communication 
is via point-to-point bidirectional links
implemented with a DropTail queue.

\noindent{\bf Scenarios.}
The scenarios are designed to test
$(i)$ the scalability of the algorithms as the number of readers, writers, and servers increases;
$(ii)$ the contention effect on efficiency, 
and 
$(iii)$ the effects of chosen topologies on the efficiency.
For scalability we test with the number of readers $|\rdSet|$ from the set $\{10, 20, 40, 80\}$
and the number of servers $|\srvSet|$ from the set $\{9, 16, 25, 36\}$.
Algorithms are evaluated with matrix quorums (unions of rows and columns). 
For the {{\small MWMR}} setting we range the number of writers $|\wSet|$
over the set $\{10, 20, 40\}$.
We issue reads (and writes) every $rInt$ (and $wInt$ respectively) from the set of $\{2,4\}$ seconds.
To test contention we define two invocation schemes:
\emph{fixed} and 
\emph{stochastic}. 
In the \emph{fixed} scheme all operations are scheduled periodically at a constant interval. 
In the \emph{stochastic} scheme reads are scheduled randomly from
the intervals $[1...rInt]$ and write operations from the intervals $[1..wInt]$.
%


%
%

\begin{figure}[thp]
{\small \centering
\begin{tabular}{c c}
		\includegraphics[width=0.46\textwidth]{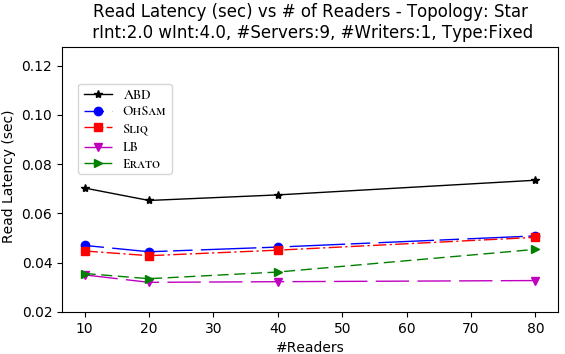}
		&
		\includegraphics[width=0.46\textwidth]{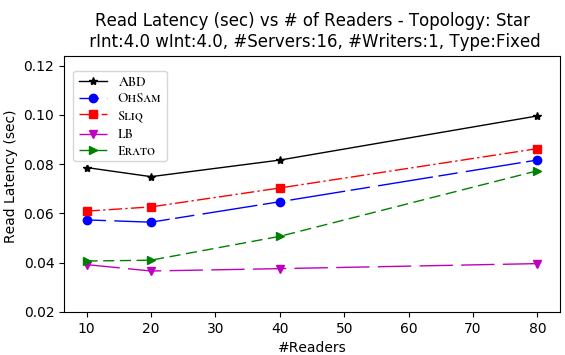} \\
		(a)  & (b) \\
		\includegraphics[width=0.46\textwidth]{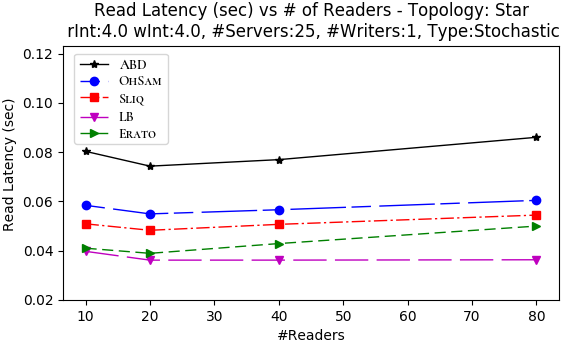} 
		&		 
		\includegraphics[width=0.46\textwidth]{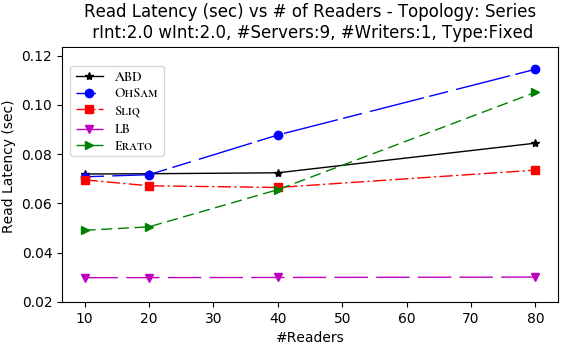}\\
			(c) & (d) \\ 
		\includegraphics[width=0.46\textwidth]{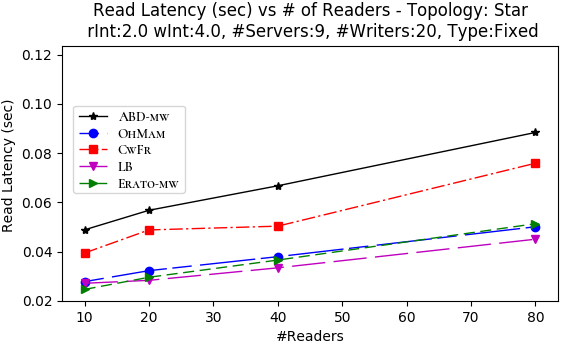}
		&
		\includegraphics[width=0.46\textwidth]{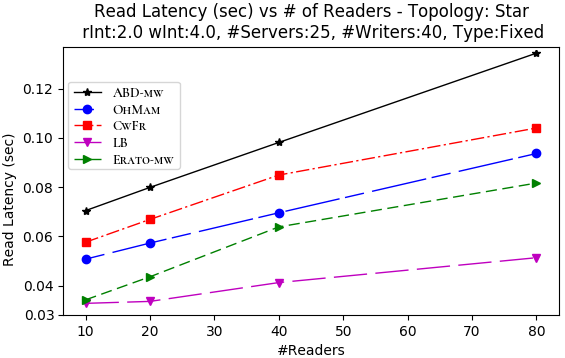} \\
		(e) & (f) \\
		\includegraphics[width=0.46\textwidth]{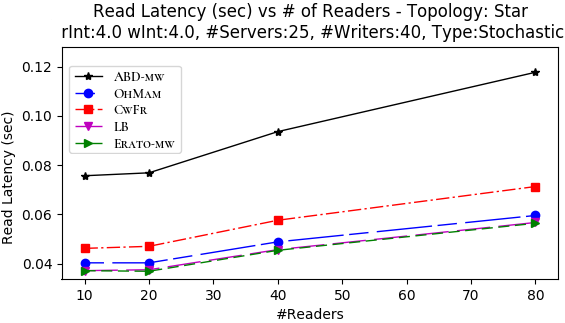} 
		&		 
		\includegraphics[width=0.46\textwidth]{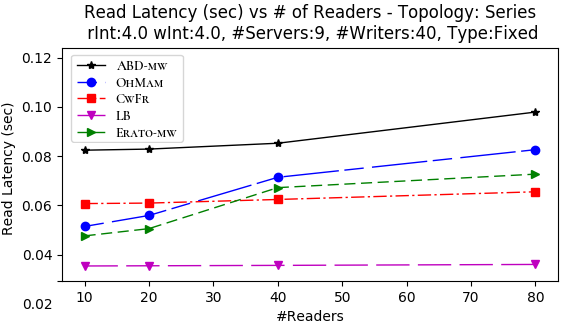}\\
				(g) & (h) \\ 
\end{tabular}
}
\vspace{-1mm}
\hspace{1cm} 
{\caption{Simulation results for  {\small SWMR} (a-d) and {\small MWMR} (e-g). 
Horizontal axis is \newline the number of readers. Vertical axis is latency.
}}
\label{fig:erato:plots}
\end{figure}

\noindent{\bf Results.}
We note that generally 
the new algorithms
outperform the competition.
A closer examination yields the following observations.

\noindent \textit{Scalability:} 
Increased number of readers and servers 
increases latency in both settings. 
Observe Fig. \ref{fig:erato:plots}(a),(b) 
 for {\small SWMR} 
and Fig. \ref{fig:erato:plots}(e),(f) 
for {\small MWMR} algorithms.
%
Not surprisingly, latency is better for smaller numbers of readers, writers, and servers.
However,  the relative performance of the
algorithms remains the same.

\noindent \textit{Contention:} 
The efficiency of the algorithms is examined under different
concurrency schemes.
%
We notice that in the \emph{stochastic} 
scheme reads complete faster 
than in the \emph{fixed} scheme -- Fig. \ref{fig:erato:plots}(b) and \ref{fig:erato:plots}(c) for the {\small SWMR}
and 
Fig. \ref{fig:erato:plots}(f) and  \ref{fig:erato:plots}(g) for the {\small MWMR} setting.
This outcome is expected as the \emph{fixed} scheme causes congestion.
For the \emph{stochastic} scheme the invocation time intervals are
distributed uniformly (randomness prevents the operations 
from being invoked simultaneously), 
and this reduces congestion in the network and improves latency.

\noindent \textit{Topology:} 
Topology substantially impacts performance and the behavior of the algorithms.
This can be seen in 
Figures \ref{fig:erato:plots}(b) and \ref{fig:erato:plots}(d) for the  {\small SWMR} setting,
and
Figures \ref{fig:erato:plots}(f) and \ref{fig:erato:plots}(h) for the  {\small MWMR} setting.
The results show clearly that the proposed algorithms outperform the competition
in the \emph{Star} topology,
where servers are well-connected using high bandwidth links.

%% file: MAIN.bbl
\begin{thebibliography}{10}

\bibitem{NS3}
{NS3} network simulator.
\newblock https://www.nsnam.org/.

\bibitem{ABD96}
{\sc Attiya, H., Bar-Noy, A., and Dolev, D.}
\newblock Sharing memory robustly in message passing systems.
\newblock {\em Journal of the ACM 42(1)\/} (1996), 124--142.

\bibitem{CDGL04}
{\sc Dutta, P., Guerraoui, R., Levy, R.~R., and Chakraborty, A.}
\newblock How fast can a distributed atomic read be?
\newblock In {\em Proceedings of the 23rd ACM symposium on Principles of
  Distributed Computing (PODC)\/} (2004), pp.~236--245.

\bibitem{EGMNS09}
{\sc Englert, B., Georgiou, C., Musial, P.~M., Nicolaou, N., and Shvartsman,
  A.~A.}
\newblock On the efficiency of atomic multi-reader, multi-writer distributed
  memory.
\newblock In {\em Proceedings 13th International Conference On Principle Of
  DIstributed Systems (OPODIS 09)\/} (2009), pp.~240--254.

\bibitem{GNRS11}
{\sc Georgiou, C., Nicolaou, N., Russel, A., and Shvartsman, A.~A.}
\newblock Towards feasible implementations of low-latency multi-writer atomic
  registers.
\newblock In {\em 10th Annual IEEE International Symposium on Network Computing
  and Applications}.

\bibitem{GNS08}
{\sc Georgiou, C., Nicolaou, N.~C., and Shvartsman, A.~A.}
\newblock On the robustness of (semi) fast quorum-based implementations of
  atomic shared memory.
\newblock In {\em DISC '08: Proceedings of the 22nd international symposium on
  Distributed Computing\/} (Berlin, Heidelberg, 2008), Springer-Verlag,
  pp.~289--304.

\bibitem{HNS2017arx}
{\sc Hadjistasi, T., Nicolaou, N., and Schwarzmann, A.~A.}
\newblock On the impossibility of one-and-a-half round atomic memory.
\newblock www.arxiv.com, 2016.

\bibitem{HNS17}
{\sc Hadjistasi, T., Nicolaou, N.~C., and Schwarzmann, A.~A.}
\newblock Oh-ram! one and a half round atomic memory.
\newblock In {\em Networked Systems - 5th International Conference, {NETYS}
  2017, Marrakech, Morocco, May 17-19, 2017, Proceedings}, pp.~117--132.

\bibitem{HW90}
{\sc Herlihy, M.~P., and Wing, J.~M.}
\newblock Linearizability: a correctness condition for concurrent objects.
\newblock {\em ACM Transactions on Programming Languages and Systems (TOPLAS)
  12}, 3 (1990), 463--492.

\bibitem{Lamport79}
{\sc Lamport, L.}
\newblock How to make a multiprocessor computer that correctly executes
  multiprocess progranm.
\newblock {\em IEEE Trans. Comput. 28}, 9 (1979), 690--691.

\bibitem{Lynch1996}
{\sc Lynch, N.}
\newblock {\em Distributed Algorithms}.
\newblock Morgan Kaufmann Publishers, 1996.

\bibitem{LS97}
{\sc Lynch, N.~A., and Shvartsman, A.~A.}
\newblock Robust emulation of shared memory using dynamic quorum-acknowledged
  broadcasts.
\newblock In {\em Proceedings of Symposium on Fault-Tolerant Computing\/}
  (1997), pp.~272--281.

\end{thebibliography}
